\documentclass[a4paper]{article}
\usepackage{amsbsy}
\usepackage{natbib}
\usepackage{graphicx}
\usepackage{subfigure}
\usepackage{bbm}
\usepackage{bm}
\usepackage{amsmath,amsthm,amssymb} 
\usepackage{color,soul}
\usepackage{dcolumn}
\usepackage{verbatim}
\usepackage[sc]{mathpazo} 
\usepackage[T1]{fontenc} 
\usepackage{microtype} 
\usepackage[hmarginratio=1:1,top=32mm,columnsep=20pt]{geometry} 
\usepackage{multicol} 
\usepackage{caption} 
\usepackage{booktabs} 
\usepackage{float} 
\usepackage{hyperref} 
\usepackage{lettrine} 
\usepackage{paralist} 
\usepackage{color,soul}
\usepackage{abstract} 
\usepackage[titletoc,toc,title]{appendix}

\usepackage{fancyhdr} 
\usepackage{todonotes}

\title{\vspace{-15mm}\fontsize{24pt}{10pt}\selectfont\textbf{Computer model calibration with large non-stationary spatial outputs: application to the calibration of a climate model}} \author{
\large
\textsc{Kai-Lan Chang and Serge Guillas}\thanks{
We thank Dr. Hanli Liu (NCAR) for technical support in running chemistry-climate model WACCM. SG gratefully acknowledges support through the NERC grant PURE (Probability, Uncertainty and Risk in the Natural Environment) NE/J017434/1.
}\\[2mm] 
\normalsize University College London, Gower Street, London WC1E 6BT, UK \\ 
\normalsize \href{mailto:ucakkac@ucl.ac.uk}{ucakkac@ucl.ac.uk} 
\vspace{-5mm}
}
\date{}

\begin{document}
\maketitle
\thispagestyle{fancy} 
\begin{abstract}
Bayesian calibration of computer models tunes unknown input parameters by comparing outputs with observations. For model outputs that are distributed over space, this becomes computationally expensive because of the output size. To overcome this challenge, we employ a basis representation of the model outputs and observations: we match these decompositions to carry out the calibration efficiently. In the second step, we incorporate the non-stationary behaviour, in terms of spatial variations of both variance and correlations, in the calibration. We insert two integrated nested Laplace approximation-stochastic partial differential equation parameters into the calibration. A synthetic example and a climate model illustration highlight the benefits of our approach.
\end{abstract}

\noindent%
{\it Keywords:} Gaussian process; integrated nested Laplace approximations-stochastic partial differential equation; Mat\'{e}rn fields; uncertainty quantification 

\section{Introduction}
\label{sec:intro}

Complex computer models are widely used in various fields of science and 
technology to mimic complex physical systems. Computer model calibration 
involves comparing the simulations of a complex computer model with the physical 
observations of the process being simulated. Increasingly, computer model 
outputs are in the form of spatial fields, particularly in environmental 
sciences. This poses a particular challenge to the calibration method. 

The class of models that we consider in this paper are computer models 
with parametric inputs of reasonable dimension (say below 20), and 
outputs distributed over two dimensions over the plane or the sphere. 
This is unlike the Kennedy and O'Hagan formulation \citep{KO01}, which 
is usually applied to scalar outputs. Our motivations come from climate 
modelling. Climate scientists compare model outputs at a certain relevant 
altitude distributed over the sphere, typically over a grid (along 
latitude and longitude), with a spatial data set of observations at 
the same altitude.

In this paper, we develop our Bayesian calibration technique based on 
the framework from \cite{KO01}: we approximate the expensive computer 
model by a Gaussian process (GP). This formulation has proven to be 
effective in a wide range of applications. However, the GP calibration 
is computationally expensive for large model output spaces (cubic 
complexity in the number of output points used to fit the GP due to 
the Cholesky decomposition). Therefore several attempts to tackle 
this issue in the context of times series of outputs or spatial outputs 
have been made either by using truncated basis representations of model 
outputs in order to reduce dimension \citep{bayarri2007computer, HGWR08, 
chang2014fast, holden2015emulation}, or by using a separable covariance 
function over space and tuning parameters to build a theoretical emulator 
for multivariate outputs \citep{rougier2008efficient, bhat2010computer}. 
We provide here a solution that makes use of an adequate representation 
of the spatial outputs using Gaussian fields.

Gaussian fields (GFs) play an important role in spatial statistics. 
The traditional approach is to specify a GF through its covariance function.  
Another approach is to use the class of Gaussian Markov random fields (GMRFs), 
which are discretely indexed GFs.  The Markov property yields a sparse precision 
matrix, so that efficient numerical algorithms can be employed. \cite{LRL11} 
show that the GMRF representation can be constructed explicitly by using a 
certain form of stochastic partial differential equation (SPDE) which has a GF 
with Mat\'{e}rn covariance as its solution. The representation employs piecewise 
linear basis functions, and Gaussian weights with Markov dependences determined 
by the finite element method over a triangulation of the domain. This technique 
can deal with large spatial data sets and naturally account for nonstationarity. 
Our paper combines the strengths of the calibration formulation with a truncated 
basis, and the SPDE defined scale and precision parameterization to deal with 
large scale spatial outputs, and still provides a compromise with computational 
feasibility in order to employ a fully Bayesian approach.

\subsection{Challenge in Bayesian calibration}

Among existing approaches of using the basis representation of model outputs, 
dimension reduction is carried out mostly by data-driven basis functions, i.e. 
principal components (PCs), also known as empirical orthogonal functions (EOFs), 
see \citet{HGWR08}. Data-driven basis functions offer a computationally 
efficient approach to adapt the outputs. For the propose of computer model 
calibration of spatial outputs, this approach ignores the nature of the spatial 
dependence of the outputs, treating spatial data as a multivariate vector. 

Since the dimension of the input space for known input parameters is two (the location in space), we could employ the usual calibration framework \citep{KO01}. However, this framework can only deal with few thousands of output points at these input locations. But climate models produce outputs over large regular grid cells, e.g. our climate model uses a grid of $n=96 \times144=13824$ cells, and this is at a rather coarse choice of resolution. We calibrate four input parameters with $r=100$ runs, thus the number of computer runs $r$, multiplied by the output size $n$ creates a data matrix that is too large to fit a GP, an impossible task for a fully Bayesian calibration (cubic complexity in the total number of output points to fit the GP). Hence our approach aims to reduce the large amount of model outputs with a smaller basis representation that makes use of the spatial dependency to extract key pieces of information, instead of using all the output cells. Our approach involves transforming a large scalar output over space into a much smaller set of scalars by using a spherical harmonics representation and the SPDE technique.

\subsection{Atmospheric chemistry model output}

We consider that an atmospheric chemistry model discretizes Earth's surface into a three-dimensional grid of cells over time, which can be characterized by horizontal (latitude and longitude), vertical (altitude or pressure level) and temporal resolutions. The output in each cell is parameterized by complex mathematical equations that describe the chemistry species in it and the physical circulation through it. The four dimensional interactions of climate dynamics are currently beyond our scope for the calibration.  Our paper only focuses on the horizontal variations. 
Our practical interest is to tune, and quantify uncertainty in, 
climate experiments. The Whole Atmosphere Community Climate Model (WACCM) is a 
general circulation model of the middle and upper atmosphere. WACCM is an 
extension of the National Center for Atmospheric Research (NCAR) Community Earth System 
Model (CESM). Many parameterizations of physical processes have to be set to run 
WACCM, resulting in potential concerns about error growth \citep{liu2009error}. 

To describe the general framework, let $\eta (\mathbf{s}_i,\boldsymbol\theta_j), 
i=1,\ldots,n; j=1,\ldots, r $ be the $r$-runs model outputs measured at $n$ 
locations. Here we refer to $m=n\times r$ as the total number of outputs in the 
simulations. We choose a design made of combinations of input values, and we 
impose distributional prior assumptions on the inputs. The 
aim of calibration is to estimate the best input setting $\boldsymbol\theta^\ast$ 
to match outputs to observations, and investigate the discrepancy between 
observations and optimized outputs. Note that in terms of the calibration framework from \cite{KO01}, our experiment does not have `known variable parameters', output cells are prescribed as a resolution in the climate model, thus the spatial variations in different model runs are completely differentiated by calibration inputs $\boldsymbol\theta$. 

For each single run, WACCM simulates output over a grid of $n=96 \times144=13824$ cells. 
We explore the zonal wind outputs over the sphere, varying according 
to four gravity wave (GW) input parameters with $r=100$ runs, in order 
to calibrate the GW parameters. The number of computer runs $r$, 
multiplied by the output size $n$ is too large to fit a GP to the 
computer model, and thus challenges the fully Bayesian calibration 
to be performed.

\subsection{The propagation of gravity waves}

In climate modeling, the GWs parametrization aims to 
reduce zonal mean wind biases. Small modification of parameterized GWs can have 
large impacts by improving the propagation pathways of the Rossby waves 
\citep{Alexander2015gw}. GWs also play a dominant role in driving the 
quasi-biennial oscillation (QBO), 
which is a dynamic process of zonal mean zonal winds from eastward to westward in the tropical stratosphere. GWs, also called small-scale atmospheric waves, generate a wide range of short horizontal wavelengths from mesoscale to thousands of kilometers \citep{ern2014interaction}, and an even wider range of the processes impacted by GWs (turbulence scales to planetary scales) \citep{liu2014gravity}. It is thus a challenge to numerically simulate all small waves and their cumulated effects that contribute to QBO pattern based on global observations \citep{alexander2010recent, geller2013comparison, yu2017sensitivity}.

\subsection{Outline of this paper}

We propose to use a fixed spatial basis, as \citet{bayarri2007computer} did 
by employing a wavelets basis to describe for functional model outputs. 
Our approach is also related to recent multi-resolution methods on spatial 
data \citep{nychka2002multiresolution, nychka2015multiresolution, ilyas2017uncertainty}. 
With a fixed basis, we can easily compare model outputs to observations over space. 
In addition, the use of a fixed basis facilitates the quantification of the 
non-stationarity across space in the SPDE model. 
 
In Section 2 we present our approach in detail. We employ a truncated basis 
representation, such as a B--splines decomposition or spherical harmonics 
transforms, to capture the output features spatially. We then explore how 
parameters in an SPDE model can explicitly quantify the nonstationarity of the 
the spatial field \citep{BL11, blangiardo2015spatial, zammit2015resolving, liu2016efficient}: we extend our approach by including spatially-varying scale and precision parameters 
in an SPDE model into our calibration framework. 
We then apply these techniques to a synthetic example in Section 3 and our real 
climate experiment in Section 4. Finally in Section 5, we discuss 
potential improvements to our approach.

\section{Methods}
\label{sec:meth}

To address the challenges of the uncertainty quantification at both global and local scales, and maintain computational feasibility for the Bayesian calibration, we pursue a sophisticated effort that approximate the spatial variations effectively and efficiently. In section 2.1, we adopt a reduced rank spatial basis representation to capture the large scale spatial variability. In the next step in Section 2.2, we review the spatial modeling technique through the SPDE approach, and highlight its strength in capturing local spatial structures. We then combine these two approaches into the calibration framework in Section 2.3, and provide a guidance for the implementation in Section 2.4.

\subsection{Basis representation for the model output}

In this section, we decompose spatial outputs and 
observations onto a basis of real valued basis functions, such as B--splines or spherical 
harmonics. We parsimoniously represent these surfaces, and construct a 
methodology for the calibration that makes use of the coefficients in these 
representations. We follow the Bayesian calibration setting of \citet{KO01}. 
Let $\boldsymbol\theta$ be the calibration parameters. The 
output $\eta(\cdot)$ is computed at inputs $(\mathbf{s},\boldsymbol\theta)$ in 
an $m$-point experimental design, where $m=n\times r$ means $r$ computer runs 
measured at $n$ locations.  The output $\eta(\mathbf{s},\boldsymbol\theta)$ is 
an approximation of the reality $y^R(\mathbf{s})$. The discrepancy between the 
simulator and the reality at the spatial locations is denoted 
$\delta(\mathbf{s})$. The observations $y^F(\mathbf{s})$ of reality are 
collected at a number of locations $\mathbf{s}$ in an $n$-point spatial design 
(here a simple grid), and are subject to a normal observation error 
$\epsilon(\mathbf{s})$ with a constant variance across locations. 
The measurement locations for observations and outputs can be different, 
since the methodology accommodates such variation. The main equation is: 
\begin{equation} \label{eq:KO}
y^F(\mathbf{s}) =y^R(\mathbf{s}) + \epsilon(\mathbf{s}) =  \eta(\mathbf{s}, 
\boldsymbol\theta^\ast) + \delta(\mathbf{s}) + \epsilon(\mathbf{s}).
\end{equation}
This formulation includes both parameter uncertainty and model discrepancy, however, it is hard to distinguish the uncertainty in the calibration parameters from discrepancy in real applications due to lack of identifiability \citep{brynjarsdottir2014learning}. Note that output cells are prescribed as a model resolution, the uncertainties in $\eta$ are completely determined by $\boldsymbol\theta$. 
We use a set of spatial basis functions  $\{ \psi_z \}$, where $z$ is an integer that 
represents the index number within the ordered basis, to decompose each run of 
model output over space. Precisely, for the $N_\eta$-th level of expansion and for each run 
$j$:

\begin{align*}
\eta(\mathbf{s},\boldsymbol\theta_j)=\sum_{z=1}^{N_\eta} c^M_{z} 
(\boldsymbol\theta_j) \psi_{z}(\mathbf{s}) \qquad 
j=1,\ldots , r.
\end{align*}
We assume that the approximation error in this representation is ignorable (i.e., we expect the more bases, the lower approximation error). The coefficients $\{c^M_z\}$ represent the surface features at different levels of expansion. Similar to \cite{nychka2015multiresolution}, we conjecture that different spatial basis functions will be valid for this representation, such as the Wendland family \citep{wendland2004scattered} used in \cite{nychka2015multiresolution} or popular spline-based approaches \citep{wood2003thin, williamson2012fast, chakraborty2013spline, bowman2016emulation, chang2017regional}. The observations can be written as (with an associated approximation error ignored):

\begin{align*}
y^F(\mathbf{s})= \sum_{z=1}^{N_y} c^F_z \psi_z (\mathbf{s}).
\end{align*}
The physical space of both model outputs and observations are 
transformed into a functional space spanned by the fixed basis.  
Since the aim is to calibrate the spatial outputs, we also assume 
that the reality $y^R(\mathbf{s})$, the discrepancy function 
$\delta(\mathbf{s})$, and the measurement errors $\epsilon(\mathbf{s})$, 
can be represented by similar basis representations, albeit with 
more levels of variation than model outputs:

\begin{align*}
y^R(\mathbf{s}) = \sum_{z=1}^{N_y} c^R_z \psi_z (\mathbf{s}), \qquad 
\delta(\mathbf{s})= \sum_{z=1}^{N_y} c^\delta_z \psi_z (\mathbf{s}),\qquad
\epsilon(\mathbf{s}) = \sum_{z=1}^{N_y} c^\epsilon_z \psi_z (\mathbf{s}).
\end{align*}
Indeed, the computer model does not include all possible physical processes that 
affect the measurements. Hence, the spatial outputs from the computer simulation 
should be relatively smoother than the observations. Therefore we assume a larger 
number of basis functions ($N_y$) in the observations (automatically as well in the discrepancy and error functions) than for model outputs ($N_\eta$, $N_\eta\leq N_y $). 
In the formulation of the calibration algorithm, we introduce coefficients 
$\{c^M_z | N_\eta < z \leq N_y \}$, all set to be $\mathbf{0}$. Indeed, we are then 
able to use the same number of basis functions $N_y$ to decompose $y^F$ and $\eta$. 
Then matching the coefficients in (\ref{eq:KO}) yields:
\begin{equation}\label{eq:linkfun}
c_{z}^F = c_{z}^R+ c^\epsilon_z=  c_{z}^M(\boldsymbol\theta^\ast) + c_{z}^\delta 
+ c^\epsilon_z, \qquad z=1,\ldots, N_y .
\end{equation}
Hence, only the relatively smooth variations of the computer model match the 
variations in observations. At this point we only seek to capture the large scale variability derived from calibration parameters, local structures will be accounted for in Section 2.2.  The weights for the measurement errors, $c^\epsilon_z$, are assumed to follow $N(0,1/\lambda_\epsilon)$. 

\subsubsection{GP for the transformed coefficients}
The GP assumption is imposed on each coefficient $c^M_{z}(\boldsymbol\theta), z=1,\ldots, N_y$, of mean 0 and with a covariance 
function
\begin{equation}\label{eq:etacov}
\text{Cov}(c^M_z(\boldsymbol\theta), c^M_{z'}(\boldsymbol\theta'))= 
\frac{1}{\lambda_\eta} I_{zz'}  \prod_{k=1}^{q} \rho_{\eta k}^{2^{\gamma_{\eta k}} |\theta_k - 
\theta'_k|^{\gamma_{\eta k}}},
\end{equation}
where $I_{zz'}$ is the Kronecker's delta ($I_{zz'}=1$ if $z=z'$ and 0 
otherwise), $q$ is the dimension of $\boldsymbol\theta$, $\lambda_\eta$ controls 
the marginal precision of $\eta(\cdot, \cdot)$ and $\boldsymbol\rho_\eta$ 
controls the strength of the dependence in each of the pairs of 
$\boldsymbol\theta$. 
To simplify the complexity and due to computer model response to input tunes is nearly smooth and continuity, it is generally reasonable to assume $\boldsymbol\gamma_\eta=2$ \citep{sacks1989design, higdon2004combining, linkletter2006variable}. Note that the coefficients $\{c^M_z\}$ have to be scaled to the unit hypercube, otherwise this covariance model is not appropriate. This reparameterization of the square exponential covariance leads to a smooth and infinitely differentiable representation for the model output \citep{stein1999interpolation}.
In addition, coefficients associated to same basis $\psi_z$ form a 
block in the covariance structure, and we assume that the correlation between 
different indices $z$ is 0. 
Hence the $rN_y$-vector $\mathbf{c}^M$ has a multivariate normal prior with mean 
0 and a covariance matrix with $r \times r$'s $N_y$ blocks in the diagonal, and 
the off-diagonal blocks are zero matrices.

The strong assumption of independence of the coefficients, through different 
blocks in the covariance, may not be fully justifiable in all real applications. Indeed, 
it is possible that certain physical properties propagate across multiple scales 
(but even in that case, it may not constitute a large proportion of the variation).
However, this assumption leads to a great computational advantage in terms of 
forming a block diagonal covariance model in the GP model. 
Traditionally a GP fitting involves a complexity of $O(m^3)=O(n^3r^3)$ and a storage cost of $O(m^2)=O(n^2r^2)$. In our approach the complexity and cost of our model are $O(N_y^3 r^3)$ and $O(N_y^2 r^2)$, where $N_y << n$. The block diagonal assumption further reduces the complexity and cost to $O(N_y r^3)$ and $O(N_y r^2)$.
In simulation study we discuss how this assumption is a compromise between fidelity and 
complexity.

The decomposed discrepancy term $c_{z}^\delta$ quantifies the inadequacy between 
the simulator and reality in the functional domain.  We assume that each $c^\delta_z$ 
follows a normal distribution of mean 0 and with a covariance function: 
\begin{equation}\label{eq:deltacov}
\text{Cov}(c^\delta_z, c^\delta_{z'})=  \frac{1}{\lambda_\delta}  I_{zz'}.
\end{equation}
There is no conceptual difference in the model bias between our setting and another setting that relies on a projection onto a basis (e.g. the PC approach), but there are differences in the ability to concretely and adequately pin down the biases. Indeed, our approach allows the bias to represent complex ranges of variations (due to its expression in a basis and the addition of nonstationarity in the sequel of this paper). Note that among existing studies identifying climate model biases, most of the biases display a systematic tendency (either underestimation or overestimation) across certain regions \citep{jun2008spatial, lamarque2013atmospheric, wang2014global, williamson2015identifying} and thus a nonstationarity feature is desirable.

All the unknown parameters in the algorithm require specified prior 
distributions which represent uncertainty about the values of these parameters. 
The following choices are made for the priors: (a) To represent our vague knowledge about calibration parameters,
we specify a uniform prior distribution over each of the calibration parameter interval; 
(b) To model the correlation parameters $\rho_{\eta_k}, k=1,\ldots, q $, a 
Beta(1, 0.1) distribution is used, which conservatively places most of its prior 
mass on values of $\boldsymbol\rho_\eta$ near 1 (indicating an insignificant 
effect); (c) Gamma prior distributions are used for each of the precision 
parameters $\lambda_\eta$, $\lambda_\delta$ and $\lambda_\epsilon$. 
Specifically, we use priors $\lambda_\eta \sim$ GAM(5, 5) (with expectation 1 
due to standardization of the responses), $\lambda_\delta \sim $ GAM(1, 0.01) 
(with expectation around 10\% of SD of the standardized responses) and 
$\lambda_\epsilon \sim $ GAM(1, 0.003) (with expectation around 5\% of SD of the 
standardized responses).

\subsubsection{The posterior distributions} 
In this stage, all the $r$-run model outputs and observations, measured over an $n$-grid cells, are reduced to transformed coefficients. 
Denote the joint $(r+1)N_y$ data vector $\mathbf{D}=(c^F, c^M)$. The sampling 
likelihood for the full data is then 
\begin{equation}\label{eq:post-lik-sh}
L(\mathbf{D}|\boldsymbol\theta,\lambda_\eta,\boldsymbol\rho_\eta, 
\lambda_\delta, \Sigma_\epsilon) \propto |\Sigma_\mathbf{D}|^{-1/2} \exp\left\{ - 
\frac{1}{2} (\mathbf{D}^T \Sigma_{\mathbf{D}}^{-1} \mathbf{D}) \right\}, 
\end{equation}
where 
\begin{align*}
\Sigma_{\mathbf{D}} = \Sigma_\eta + \begin{pmatrix}
\Sigma_\epsilon + \Sigma_\delta & 0 \\
0 & 0 \\
\end{pmatrix},
\end{align*}
in which $\Sigma_\epsilon$ is the $N_y \times N_y$ observation covariance 
matrix, $\Sigma_\eta$ is obtained for each pair of $(r+1)N_y$ simulation inputs 
through (\ref{eq:etacov}) corresponding to $\mathbf{D}$, and $\Sigma_\delta$ is 
an $N_y \times N_y$ matrix obtained for each pair of $N_y$ input through the 
instances of (\ref{eq:deltacov}) that correspond to the coefficients $c^F$.
Let $\pi(\boldsymbol\theta)$ be the joint prior distribution for the (unknown) 
calibration vector $\boldsymbol\theta$. The resulting posterior density has the 
form 
\begin{equation}\label{eq:post-dis-sh}
\pi(\boldsymbol\theta ,\lambda_\eta,\boldsymbol\rho_\eta, \lambda_\delta 
|\mathbf{D})
\propto L(\mathbf{D}|\boldsymbol\theta,\lambda_\eta,\boldsymbol\rho_\eta, 
\lambda_\delta, \Sigma_\epsilon)\times \pi(\boldsymbol\theta) \times \pi(\lambda_\eta)\times 
\pi(\boldsymbol\rho_\eta)\times \pi(\lambda_\delta),
\end{equation}
which can be explored via a Markov chain Monte Carlo (MCMC) technique, for which 
we employ a Metropolis--Hastings algorithm. The calibrated vector is then 
denoted by $\boldsymbol\theta^\ast=\text{argmax}_\theta \pi(\boldsymbol\theta 
,\lambda_\eta,\boldsymbol\rho_\eta, \lambda_\delta |\mathbf{D})$. We 
implement a Metropolis--Hastings algorithm to produce the realization of the posterior.
  Metropolis updates are used for the correlation and the 
calibration parameters with a uniform proposal distribution centered at the 
current value of the parameter. The precision parameters are sampled using 
Hastings updates with a uniform proposal distribution centered at the current 
value of the parameter \citep{HGWR08}. This eventually yields draws from the 
posterior distribution by repeatedly accepting and rejecting a choice of move in 
the parameter space. 

The specification of the covariance structures for the truncated basis representation is a
mathematical challenge: finding explicit expressions for the covariance is hard
\citep{JS08}.  There is an alternative way to efficiently model 
complex spatial covariance structures with the added bonus of a suitable 
depiction of the nonstationarity structure into our calibration algorithm: the SPDE approach.  We introduce it in the next section.

\subsection{Spatial modeling through the SPDE approach}

Traditional models in spatial statistics build an approximation of the 
entire underlying random field. They are usually specified through the covariance 
function of the latent field. In order to assess uncertainties in the spatial 
interpolation over the whole spatial domain, we cannot build models only for the 
discretely located observations or  model outputs, we need to build an 
approximation of the entire underlying stochastic process defined on the spatial 
field.  We consider statistical models for which the unknown functions are 
assumed to be realizations of a Gaussian random spatial process. The conventional 
fitting approach spatially interpolates values as linear combinations of the 
original observed locations, and this constitutes the spatial kriging predictor. 

Due to the fixed underlying covariance structure, this approach requires more 
sophisticated treatments to take into consideration nonstationarity 
\citep{stein2005space, JS08, yue2010nonstationary, kleiber2012nonstationary, gramacy2015local}. 
A different computational approach was introduced by \cite{LRL11}, in which 
random fields are expressed as a weak solution to an SPDE, with explicit links 
between the parameters of the SPDE model and the Mat\'{e}rn covariance function. 
In this section we review some of the main concepts in spatial modeling through 
the SPDE approach. 

It may seem contradictory to make use of the SPDE approach since it seemingly only captures local structures, and climate model outputs display smooth variations. However, the SPDE approach, especially the nonstationarity version, is able to translate these smooth variations of the model outputs (and of observations) into a statistical description of the variations across space that efficiently characterizes the spatial behaviour (through the scale and precision parameters). Spatially distributed observations will still display more erratic behaviour than model outputs, but the SPDE approach will allow the calibration to be steered only by the parameters associated with the smoothest components.

\subsubsection{Mat\'{e}rn covariance and the link to SPDE} 
The Mat\'{e}rn function is a flexible covariance structure and widely used in the spatial statistics \citep{stein2005space, jun2007approach, JS08, gneiting2010matern, genton2015cross}. The choice of covariance is not that important indeed for calibration parameters \citep{KO01}, but for the outputs (across location inputs), the choice of covariance is essential, as we show. 
The shape parameter $\nu>0$, the scale parameter $\kappa>0$, and the marginal precision $\tau^2>0$, parameterize it:

\begin{equation*}
\text{Cov}(\mathbf{h})=\frac{2^{1-\nu}}{(4\pi)^{d/2}\Gamma(\nu 
+d/2)\kappa^{2\nu} \tau^2} (\kappa\|\mathbf{h}\|)^\nu K_\nu 
(\kappa\|\mathbf{h}\|), \mathbf{h} \in \mathbb{R}^d,
\end{equation*}
where $\mathbf{h}$ denotes the difference between any two locations $s$ and 
$s'$: $\mathbf{h}=s-s'$, and $K_\nu$ is the modified Bessel function of the 
second kind of order $\nu$. 

We denote by $Y(\mathbf{s})$ the observations (or the spatially distributed 
model outputs) for a latent spatial field $X(\mathbf{s})$, with a Mat\'{e}rn 
covariance structure. We assume a zero mean Gaussian noise, 
$\mathcal{W}(\mathbf{s})$, with a constant variance $\sigma^2_s$: $Y(\mathbf{s}) 
= X(\mathbf{s}) + \mathcal{W}(\mathbf{s})$. Thus, according to \cite{Whittle63}, 
the latent field $X(\mathbf{s})$ is the solution of a stationary SPDE:

\begin{equation}\label{eq:matern}
(\kappa^2 - \Delta)^{\alpha/2} \tau X(\mathbf{s})=\mathcal{W}(\mathbf{s}),
\end{equation} 
where $\Delta$ is the Laplace operator. We explain in the next paragraph how the 
analysis of this SPDE can be carried out by the finite element method.  
The regularity (or smoothness) parameter $\nu$ essentially determines the order 
of differentiability of the fields. The link between the Mat\'{e}rn field and 
the SPDE is given by $\alpha=\nu +d/2$, which makes explicit the relationship 
between dimension and regularity for fixed $\alpha$. On more general manifolds 
than $\mathbb{R}^d$, such as the sphere \citep{amt-8-4487-2015}, the direct Mat\'{e}rn representation is not easy to implement (for example, Mat\'{e}rn covariance 
with great circle distance is only valid at $\nu \in (0, 0.5]$ 
\citep{gneiting2013strictly}), but the SPDE formulation provides a natural 
generalization, and the $\nu$-parameter will keep its meaning as the 
quantitative measure of regularity.  Instead of defining Mat\'{e}rn fields by 
the covariance function, \cite{LRL11} used the solution of the SPDE as a 
definition, and it is much easier and flexible to do so. This definition also 
facilitates nonstationary extensions by allowing the SPDE parameters $\kappa$ 
and $\tau$ in Eq. (\ref{eq:matern}) to vary with space, hence denoted 
$\kappa(\cdot)$ and $\tau(\cdot)$ respectively.

\subsubsection{SPDE model construction} 
We estimate the SPDE parameters and supply uncertainty information about the 
spatial fields by using the \textit{integrated nested Laplace approximations} 
(INLA) framework, available as an R package (\url{http://www.r-inla.org/}) 
\citep{lindgren2015bayesian, rue2017bayesian}. 
The models implemented in the INLA--SPDE framework 
are built on a basis representation (triangulation over the spatial domain): 
$X(\mathbf{s}) = \sum_{i=1}^{M} \varphi_i(\mathbf{s})w_i$, 
where $\{w_i\}$ are the stochastic weights chosen so that the distribution of 
the functions $X(\mathbf{s})$ approximates the distribution of solutions to the 
SPDE on the space, and $\varphi_i(\mathbf{s})$ are piecewise linear basis with 
compact support (i.e. finite elements) in order to obtain a Markov structure, 
and to preserve it when conditioning on local observed locations. The Markov 
property yields a sparse precision matrix, so that efficient numerical 
algorithms can be employed for large spatial data. The projection of the SPDE 
onto the basis representation is chosen by a finite element method. The finite 
element method represents a general class of techniques for the approximate 
solution to partial differential equations. The 
piecewise linear basis functions defined by a triangulation of the spatial 
domain allow us to explicitly evaluate the precision matrix of the latent field. 
As a result, $X(\mathbf{s})$ follows a normal distribution with mean 0, and the 
precision matrix can be explicitly expressed as a combination of the piecewice 
linear basis functions weighted by $\kappa$ and $\tau$ (which means $\kappa$ and 
$\tau$ have a joint influence on the marginal variances of the latent field). 
Then $X(\mathbf{s})$ can be generated continuously as approximative solutions to 
the SPDE.

For the WACCM output domain, the triangulation is simply built upon regularly 
gridded cells. Note that the triangulation can be made adaptive to the 
irregularly distributed spatial data \citep{cameletti2013spatio}.  
The default value in INLA is $\alpha = 2$, but $0\leq \alpha<2$ are also 
available, though yet to be completely tested \citep{lindgren2015bayesian}. 
So, with a 2-dimensional manifold (e.g. $\mathbb{R}^2$ and $\mathbb{S}^2$), 
the smoothness parameter $\nu$ must be fixed at 1 due to the relationship 
$\alpha=\nu +d/2$. The strength of this SPDE technique enables us to 
quantify the level of nonstationarity by employing spatial basis 
representations for both $\kappa$ and $\tau$ (i.e. these quantities are 
constants in a stationary field). With a focus on the calibration, let 
$\kappa^M(\mathbf{s}, \boldsymbol\theta)$ and $\tau^M(\mathbf{s}, 
\boldsymbol\theta)$ be the scale and precision parameters in an SPDE 
model used to approximate the model outputs. To obtain basic identifiability, 
$\kappa^M(\mathbf{s}, \boldsymbol\theta)$ and $\tau^M(\mathbf{s}, 
\boldsymbol\theta)$ are taken to be positive, and their logarithm can 
be decomposed as:

\begin{equation*}
\log\kappa^M(\mathbf{s}, \boldsymbol\theta_j) = \sum_{z=1}^{N_\kappa} 
\kappa^M_{z}(\boldsymbol\theta_j) \psi_{z}(\mathbf{s}), \quad \text{and}  \quad
\log\tau^M(\mathbf{s},  \boldsymbol\theta_j) = \sum_{z=1}^{N_\tau} 
\tau^M_{z}(\boldsymbol\theta_j) \psi_{z}(\mathbf{s}), 
\quad j=1,\ldots, r.
\end{equation*}
Each basis function is evaluated at output cells and observed locations. 
The coefficients $\{ \kappa^M_z \}$ and $\{ \tau^M_z \}$ represent local 
variances and correlation ranges \citep{BL11, LRL11, fuglstad2015exploring}. 
For the sake of simplicity, we call these coefficients `SPDE parameters' in 
the calibration. We introduce in next section how to incorporate the SPDE 
parameters into calibration in order to enhance the prediction accuracy.

\subsection{Combining SPDE modeling and calibration}
A reduced rank approach was often used to ease the computational issue in large spatial data sets \citep{banerjee2008gaussian, cressie2008fixed, furrer2009spatial, katzfuss2011spatio}. In order to reduce and summarize a spatial field properly, both global and local scale dependences need to be well captured and represented. To do so, a two steps approximation was developed by combining the reduced rank representation and sparse matrix techniques, to account for global and local structures, respectively \citep{Stein07, sang2012full}. We follow the same idea of using a reduced rank representation to capture global scale variability (described in Section 2.1), while instead of tapering the covariance matrix into sparse, we use the INLA-SPDE technique to represent small scale variability. 
In this section we describe the details of our extension by including the 
SPDE defined scale and precision parameters into the Bayesian calibration.

As $\{\kappa^M_z(\boldsymbol\theta)\}$ and $\{\tau^M_z(\boldsymbol\theta)\}$ can 
quantify the nonstationarity and derivative information in the spatial process, 
we now include these two types of coefficients into our technique (combined with 
$\{c^M_z(\boldsymbol\theta)\}$ in the previous section, and vectorized all 
coefficients as a scalar). Then our approach represents the observations and 
model input--output relationship as follows:

\begin{align*}
y^F(s_1),\ldots, y^F(s_n) &\xrightarrow{\text{transform}} c^F_1,\ldots, c^F_{N_y}, 
\kappa^F_1, \ldots, \kappa^F_{N_\kappa}, \tau^F_1,\ldots, \tau^F_{N_\tau} \\
\eta(s_1, \theta_1),\ldots, \eta(s_n, \theta_1) &\xrightarrow{\text{transform}} 
c_1^M(\theta_1),\ldots, c^M_{N_y}(\theta_1),
\kappa^M_1(\theta_1), \ldots, \kappa^M_{N_\kappa}(\theta_1), \tau^M_1(\theta_1),\ldots, \tau^M_{N_\tau}(\theta_1) \\
& \vdots  \\
\eta(s_1, \theta_r),\ldots, \eta(s_n, \theta_r) &\xrightarrow{\text{transform}} 
c_1^M(\theta_r),\ldots, c^M_{N_y}(\theta_r), 
\kappa^M_1(\theta_r), \ldots, \kappa^M_{N_\kappa}(\theta_r), \tau^M_1(\theta_r),\ldots, \tau^M_{N_\tau}(\theta_r) \\
\end{align*}

where $N_y + N_\kappa+N_\tau <<n$. 
The aim is to combine the SPDE parameters as nonstationary 
information for the implementation of the calibration algorithm, and 
to model all coefficients jointly with the GP assumption.
 We also assume that the three types of coefficients are 
independent.  To describe the formulation of the design matrix, let $\{z_1, z_2, 
z_3 | z_1=1,\ldots, N_y; z_2=1,\ldots, N_\kappa; z_3=1,\ldots, N_\tau \}$ be the
indices used to represent each triplet of coefficients, 
respectively. The calibration formulation is hence:

\begin{align*}
\begin{pmatrix}
c^F_{z_1} \\
\kappa^F_{z_2} \\
\tau^F_{z_3}
\end{pmatrix}
= 
\begin{pmatrix}
c^M_{z_1}(\boldsymbol\theta) \\
\kappa^M_{z_2}(\boldsymbol\theta) \\
\tau^M_{z_3}(\boldsymbol\theta)
\end{pmatrix}
+
\begin{pmatrix}
c^\delta_{z_1} \\
\kappa^\delta_{z_2} \\
\tau^\delta_{z_3}
\end{pmatrix}
+
\begin{pmatrix}
c^\epsilon_{z_1} \\
\kappa^\epsilon_{z_2} \\
\tau^\epsilon_{z_3}
\end{pmatrix}.
\end{align*}

Thus there are $(N_y + N_\kappa + N_\tau)$-blocks of coefficients corresponding 
to each combination of $\boldsymbol\theta_j, j=1\ldots, r$ in the covariance 
matrix. The GP assumption is imposed on each coefficient $(c^M_{z_1,j}, 
\kappa^M_{z_2,j}, \tau^M_{z_3,j} )^T$ with mean 0 and covariance function

\begin{align*}
\text{Cov} ((c^M_{z_1}(\boldsymbol\theta), 
\kappa^M_{z_2}(\boldsymbol\theta),\tau^M_{z_3}(\boldsymbol\theta))^T, 
(c^M_{z'_1}(\boldsymbol\theta'), 
\kappa^M_{z'_2}(\boldsymbol\theta'),\tau^M_{z'_3}(\boldsymbol\theta'))^T)
=\frac{1}{\lambda_\eta} \prod_{i=1}^3 I_{z_i z'_i} \prod_{k=1}^q \rho_{\eta 
k}^{4(\theta_k - \theta'_k)^2},
\end{align*}

where $I_{z_i z'_i}=1$ if $z_i=z'_ i$ and 0 otherwise.
In other words, these 3 types of coefficients $\{{c}^M_{z_1}, \kappa^M_{z_2}, 
\tau^M_{z_3} \}$ have a joint multivariate normal prior distribution with mean 
0, and a covariance structure forming a block diagonal matrix:

\begin{align*}
\begin{pmatrix}
c^M_{z_1} \\
\kappa^M_{z_2} \\
\tau^M_{z_3}
\end{pmatrix}
\sim N \left(\mathbf{0}, \begin{pmatrix}
\text{Cov}(c^M_{z_1}(\boldsymbol\theta), c^M_{z'_1}(\boldsymbol\theta')) & 0 & 0 
\\
0 & \text{Cov}(\kappa^M_{z_2}(\boldsymbol\theta), 
\kappa^M_{z'_2}(\boldsymbol\theta')) & 0 \\
0 & 0 & \text{Cov}(\tau^M_{z_3}(\boldsymbol\theta), 
\tau^M_{z'_3}(\boldsymbol\theta')) \\
\end{pmatrix}
\right).
\end{align*}

The elements in each block are also block diagonal matrices. The model 
discrepancy term in the functional space follows a GP assumption defined in 
Equation (\ref{eq:deltacov}).  All the prior assumptions discussed in the 
previous section remain unchanged.  Thus the sampling likelihood in 
(\ref{eq:post-lik-sh}) and the posterior distribution in (\ref{eq:post-dis-sh}) 
still hold in this case. Overall, we decompose the model outputs into a basis 
via the coefficients $\{c^M\}$, and estimate the SPDE parameters $\{ \kappa^M, 
\tau^M \}$ in the latent field through a regression onto these basis functions. 
We are essentially fitting a GP model with $\{c^M\}$ for the regression mean 
structure and $\{ \kappa^M, \tau^M \} $ for the parameters of Mat\'{e}rn 
covariance function.

\subsection{Guidance for the number of basis functions}
In real applications, we often do not know whether the calibrated values work until 
actually performing a validation. It can be computationally challenging to find 
the optimized orders for the combination of $N_y$, $N_\kappa$ and $N_\tau$. 
Similar to most truncated basis representations, we choose the number of basis 
functions \textit{post hoc}. We provide the following model selection guidelines:  
(1) The basis representation for the mean structure of model outputs play a dominant 
role in the algorithm. Typically we cannot expect to calibrate a global process 
only through a local structure. Therefore $N_y$ usually needs to be greater 
than $N_\kappa + N_\tau$; (2) Calibration with only one of the coefficients 
$\kappa$ or $\tau$ cannot improve the analysis. The reason is the fact that 
$\kappa$ and $\tau$ represent a spatial process jointly being tacitly assumed.  
Recall that the Mat\'{e}rn function is controlled by the smoothness parameter 
$\nu$, the scale parameter $\kappa$, and the precision parameter $\tau$.  
The parameter $\nu$ is fixed by $\alpha=\nu+d/2$ in connection with the SPDE, 
thus the approximated spatial process depends upon $\kappa$ and $\tau$ jointly.  
Both $\kappa$ and $\tau$ need to be included to reflect the full variation in 
the spatial field. 

In this paper we use spherical harmonics (SHs) as our primary investigation. 
The SHs represent the wave features at different scales on the sphere \citep{BL11, JS08}. 
For the purpose of calibration, it seems unnecessary in general to approximate the 
spatial processes with very high order expansions of SHs to fit each run 
of model output best. The main requirement is to extract sufficient and meaningful 
information about the calibration parameters from the variations in the SH 
coefficients that could be attributed to variations in the inputs. To ensure 
that this requirement is met, a simple validation is to increase the basis 
number and re-calibrate the model. In case the results have no statistically 
significant impacts, then the number is large enough. \citet{muir2015method} 
utilized the corrected Akaike information criterion (AIC) to choose an optimal 
maximum order of expansion for an irregular data on the sphere in a hierarchical 
Bayesian setting. The results show the 3rd to 5th order of expansion in SHs are 
generally a turning point from fast to slow reduction in AIC in terms of balancing 
explanatory power with simplicity (although not the smallest AIC). In all these 
approaches, the choice of the number of basis vectors is currently \textit{post 
hoc}. We reckon that the 3rd or 4th order of SH transform for capturing large scale 
variability, along with a lower order of SPDE nonstationary information to account 
for local structure, as a good start in practical application.

\section{Simulation Study: Nonstationary field}
\label{sec:verify}

\begin{figure}[bht!]
    \begin{center}
        \subfigure[Combined surface]{%
            \includegraphics[height=4cm, width=0.5\textwidth]{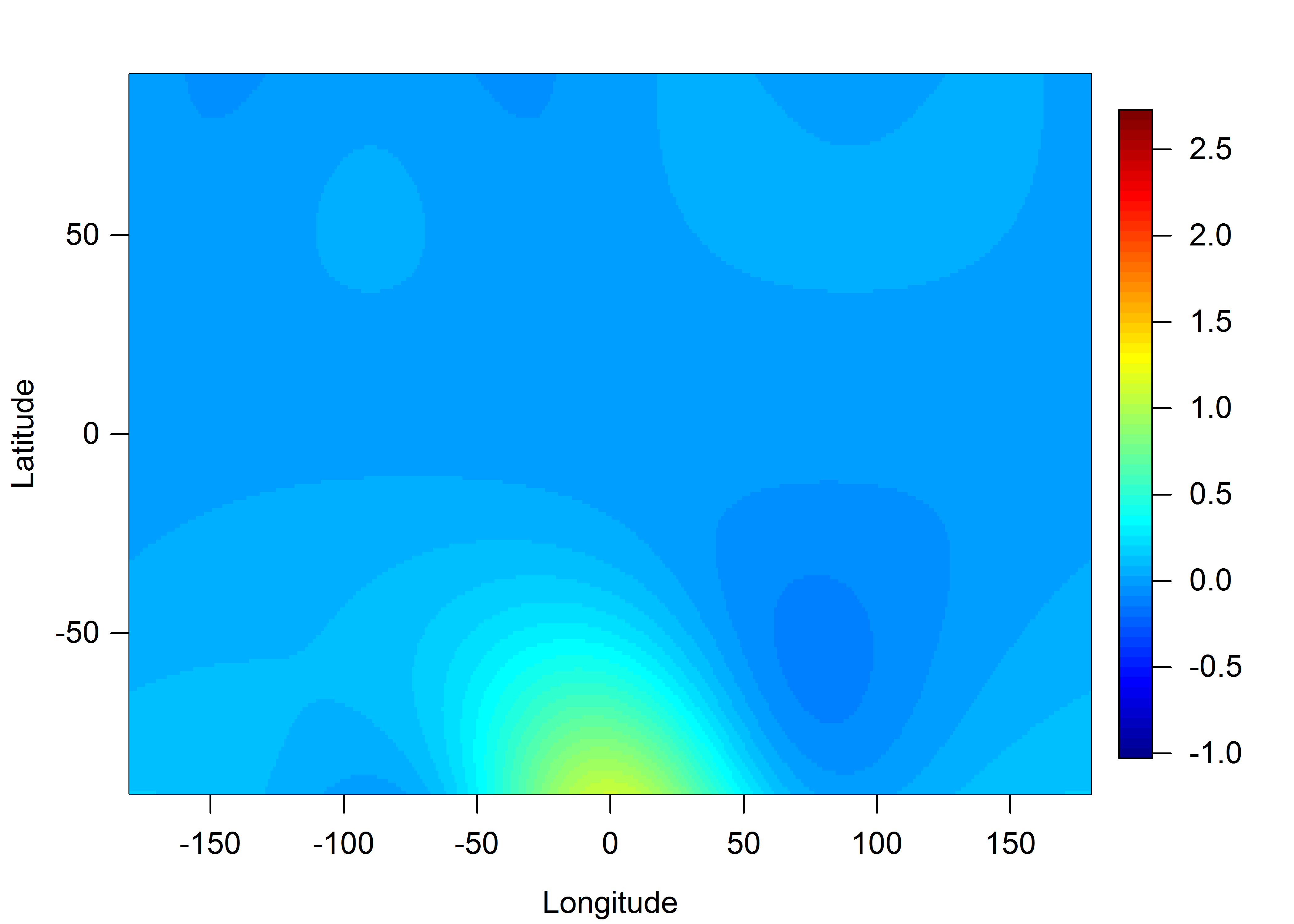}
        }%
        \subfigure[$s_2 s_3$]{%
            \includegraphics[height=4cm, width=0.5\textwidth]{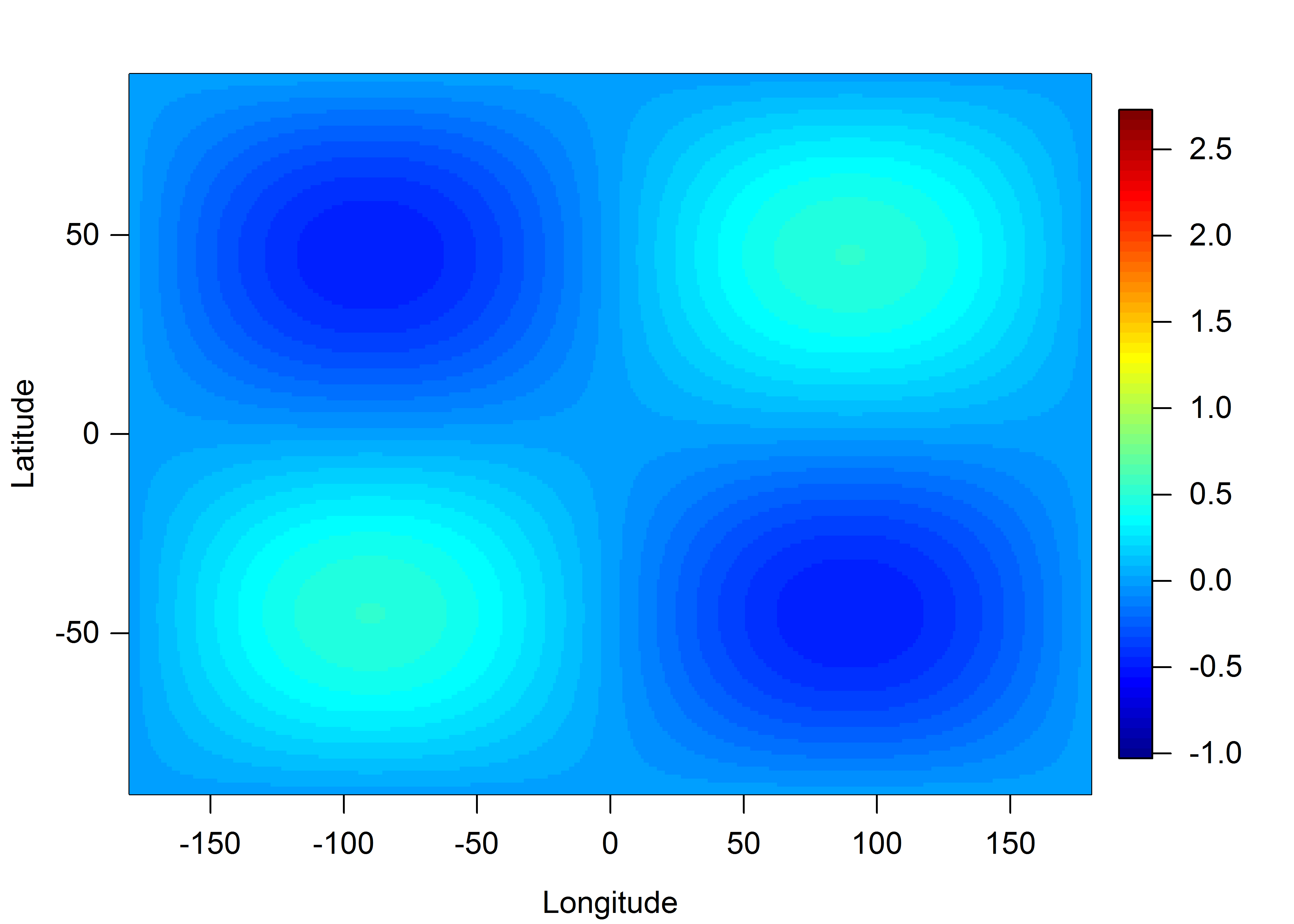}
        }\\  
        \subfigure[$s_2$]{%
            \includegraphics[height=4cm, width=0.5\textwidth]{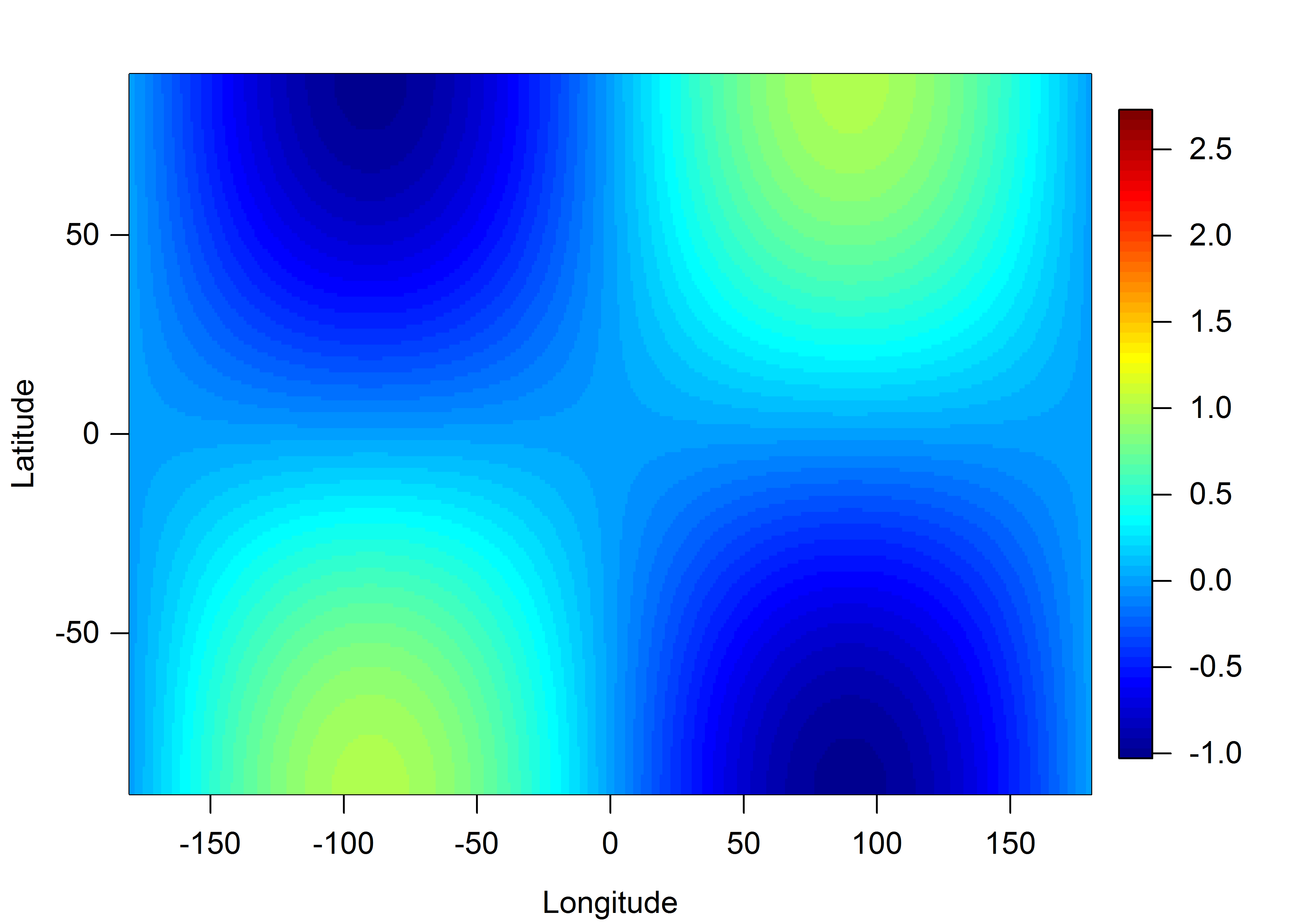}
        }%
        \subfigure[$\exp(-s_3 - s_1)$]{%
            \includegraphics[height=4cm, width=0.5\textwidth]{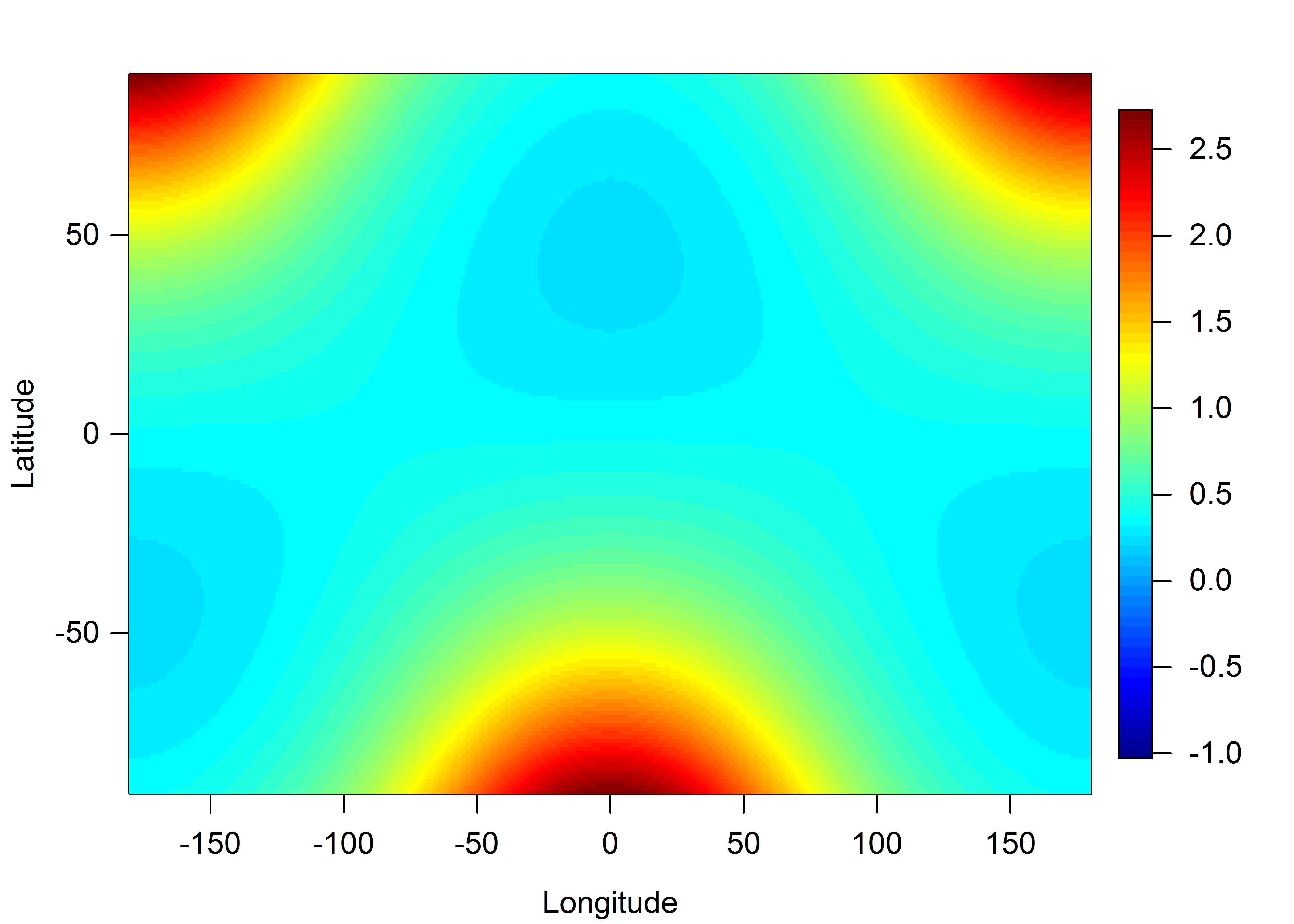}
        }\\          
    \end{center}
    \caption{Combined synthetic observed surface and each spatial component in function (\ref{eq:ex5}).}
    \label{fig:ex5-surface}
\end{figure}

In order to illustrate the methodology, this synthetic example 
simulates a nonstationary field on the sphere, with an anisotropic property (the 
spatial correlation depends on latitude), to demonstrate how including the 
parameters in the SPDE can enhance the GP calibration in such situations. 
We illustrate how the parameters in an SPDE technique can be incorporated 
into our calibration algorithm to model nonstationarity over a spherical domain. 
With $n=10\times 10$ regularly spaced locations in latitude ($L$) and 
longitude ($l$), and $r=50$ computer runs according to a maximin latin 
hypercube design (LHD) for the calibration inputs, the 
function with three calibration parameters ($q=3$) is set to
\begin{equation}\label{eq:ex5}
f(\tilde{\mathbf{s}},\boldsymbol\theta) = (0.5 s_1^2  + \theta_1 s_2 
s_3) \times \left\{ 
  \begin{array}{l l}
    \theta_2 s_2  & \quad \text{if $L > \pi/2$}\\
    \theta_3 \exp(-s_3 - s_1) & \quad \text{if $L \leq \pi/2$}
  \end{array} \right., (\theta_1, \theta_2, \theta_3)\in [0,1]^3,
\end{equation}
\sloppy where the true values for $(\theta_1, \theta_2, \theta_3)$ are set to $(0.5, 
0.2, 0.8)$, and $(s_1, s_2, s_3)=(\cos(l)\sin(L), \sin(l)\sin(L), \cos(L))$ are spherical coordinates. 
We create a nonstationary spatial field by introducing different 
structures in the Northern and Southern hemispheres, where $\theta_1$ is a 
global calibration parameter, and $(\theta_2, \theta_3)$ are local variates. 
Note that in this example the local structures are designed to be larger than global structure: $\exp(-s_3 - s_1)$ has stronger variation than $s_2$, and both of them have a larger variation than $s_2 s_3$ (see the magnitude of variation in each component from Fig. \ref{fig:ex5-surface}).

First, we perform the spherical harmonics transform (SHT) onto observations $y^F$ 
and each computer run $\eta_j, j=1,\ldots, 50$, and then carry out the 
calibration on the coefficients. In total, we estimate 13 models with different 
numbers of expansion order. The results of using the 4th to 7th orders of the 
SHT are shown in the first part of Table \ref{tab:ex5-post} (Strategies A-D).  
We can see that the global calibration parameter $\theta_1$ is estimated well. 
However, even though the convergence of an MCMC chain can be established 
for $\theta_2$ and $\theta_3$, the posterior means are underestimated. 
According to the root-mean-square error (RMSE) between assumed and predicted 
observations, which can be written as 
\begin{align*}
\text{RMSE} = \sqrt{\frac{\sum_{i=1}^{n} \left(f(s_i, \boldsymbol\theta^\ast) - f(s_i, \boldsymbol\theta^{post}) \right)^2}{n}}, 
\end{align*} 
an increase in the expansion order cannot improve the results. This 
underestimation can be viewed as a deficiency to capture local variations 
through a global mean structure: the variation created by these two 
parameters will be obscured and distorted by the variation from $\theta_1$.
 
\begin{table}
\caption{\label{tab:ex5-post}Posterior mean and SD for $(\theta_1, \theta_2, 
\theta_3 )$ in function (\ref{eq:ex5}), RMSE and number of coefficients (right 
column) under different orders of SHT for $\{\eta, \kappa, \tau\}$ per model 
run (RMSE was multiplied by $10^3$ to illustrate the magnitude).}
\centering
\begin{tabular}{c c c c r r r r | c}
\hline
Strategy & $\eta$ & $\kappa$ & $\tau$ & $\theta_1$(=0.5) & $\theta_2$(=0.2) & 
$\theta_3$(=0.8) & RMSE & $N_y +N_\kappa + N_\tau$ \\
\hline
A & 4 & - & - & 0.505(0.050) & 0.188(0.048) & 0.762(0.038) & 92 & 15\\
B & 5 & - & - & 0.498(0.053) & 0.179(0.062) & 0.746(0.050) & 132 & 21\\
C & 6 & - & - & 0.477(0.062) & 0.166(0.079) & 0.705(0.069) & 237 & 28\\
D & 7 & - & - & 0.488(0.112) & 0.198(0.127) & 0.695(0.119) & 257 & 36\\
\hline
E & - & 1 & 1 & 0.579(0.158) & 0.148(0.068) & 0.620(0.200) & 431 & 6\\
F & - & 2 & 2 & 0.560(0.097) & 0.189(0.078) & 0.740(0.089) & 145 & 12\\
G & - & 3 & 3 & 0.785(0.078) & 0.442(0.155) & 0.858(0.054) & 433 & 20\\
\hline\
H & 4 & 1 & 1 & 0.452(0.097) & 0.071(0.049) & 0.495(0.037) & 755 & 21\\
I & 5 & 1 & 1 & 0.495(0.044) & 0.133(0.049) & 0.498(0.032) & 737 & 27\\
J & 6 & 1 & 1 & 0.356(0.050) & 0.135(0.052) & 0.686(0.119) & 322 & 34\\
K & 4 & 2 & 2 & 0.553(0.068) & 0.225(0.108) & 0.771(0.109) & 80 & 27\\
L & 5 & 2 & 2 & 0.529(0.068) & 0.179(0.107) & 0.794(0.098) & 28 & 33\\
M & 6 & 2 & 2 & 0.537(0.066) & 0.171(0.110) & 0.789(0.083) & 39 & 40\\
\hline
\end{tabular}
\end{table}

\begin{figure}[bth]
    \begin{center}
        \subfigure[1st order $\kappa^F$]{%
            \includegraphics[height=4cm, width=0.5\textwidth]{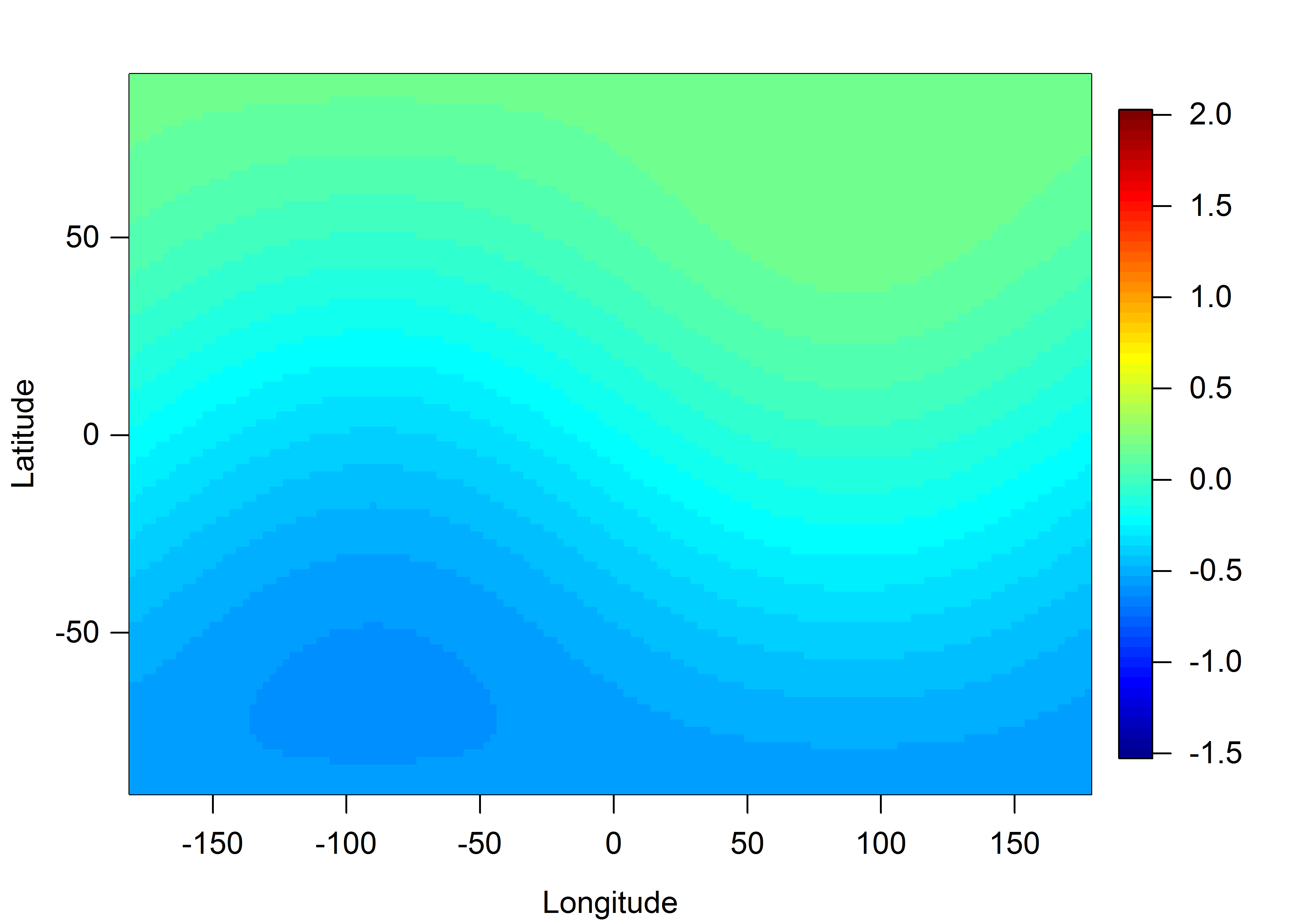}
        }%
        \subfigure[1st order $\tau^F$]{%
            \includegraphics[height=4cm, width=0.5\textwidth]{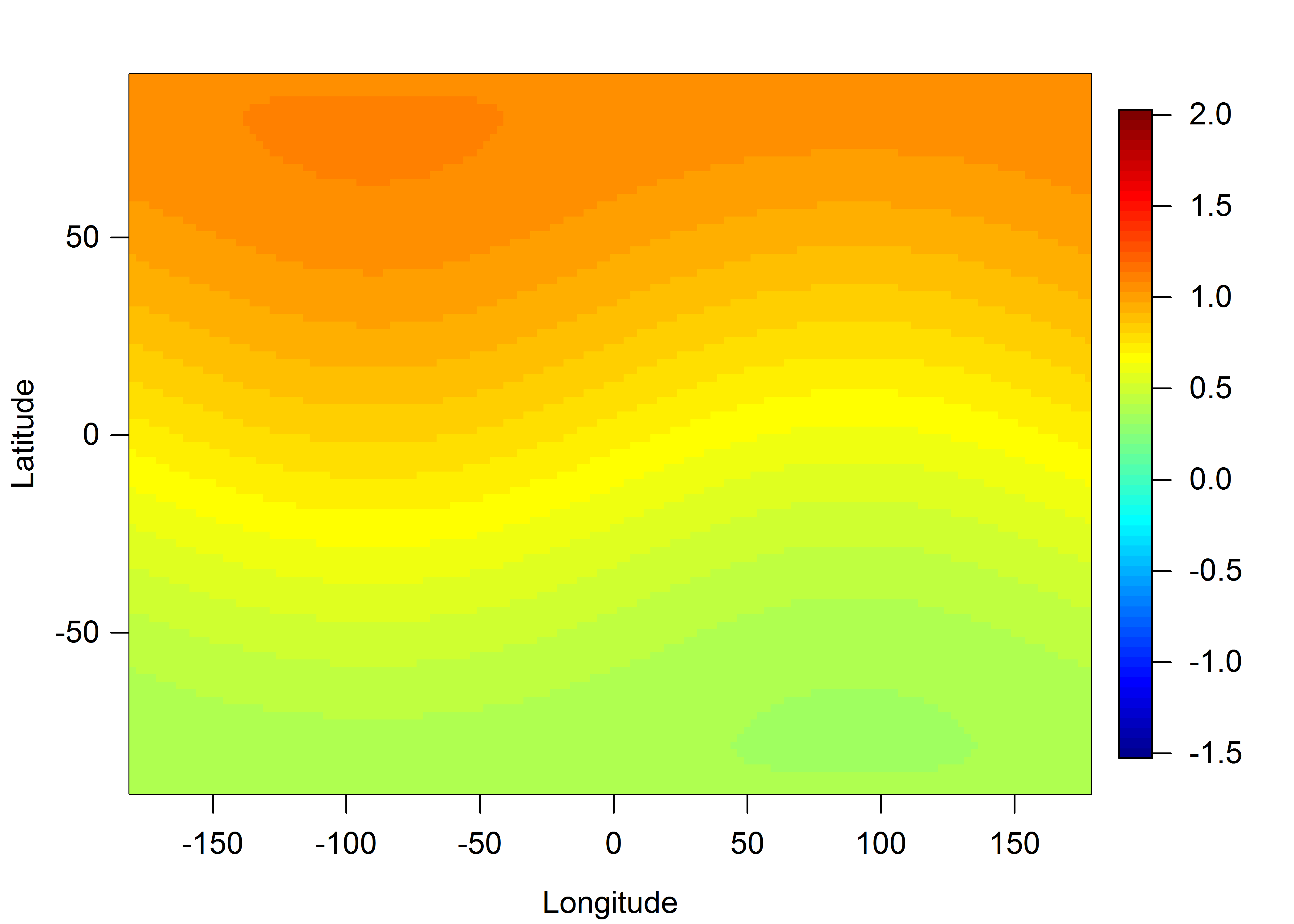}
        }\\  
        \subfigure[2nd order $\kappa^F$]{%
            \includegraphics[height=4cm, width=0.5\textwidth]{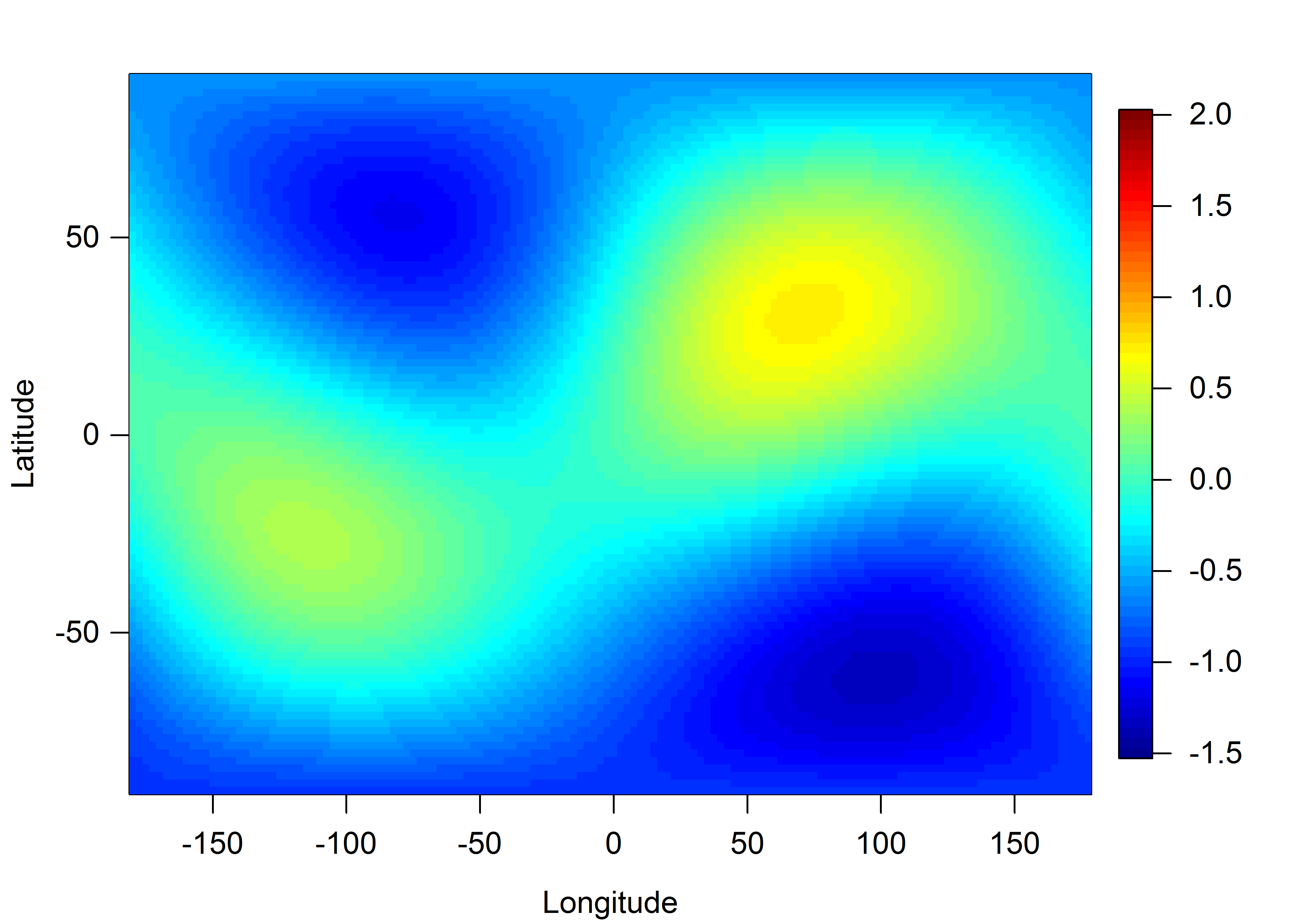}
        }%
        \subfigure[2nd order $\tau^F$]{%
            \includegraphics[height=4cm, width=0.5\textwidth]{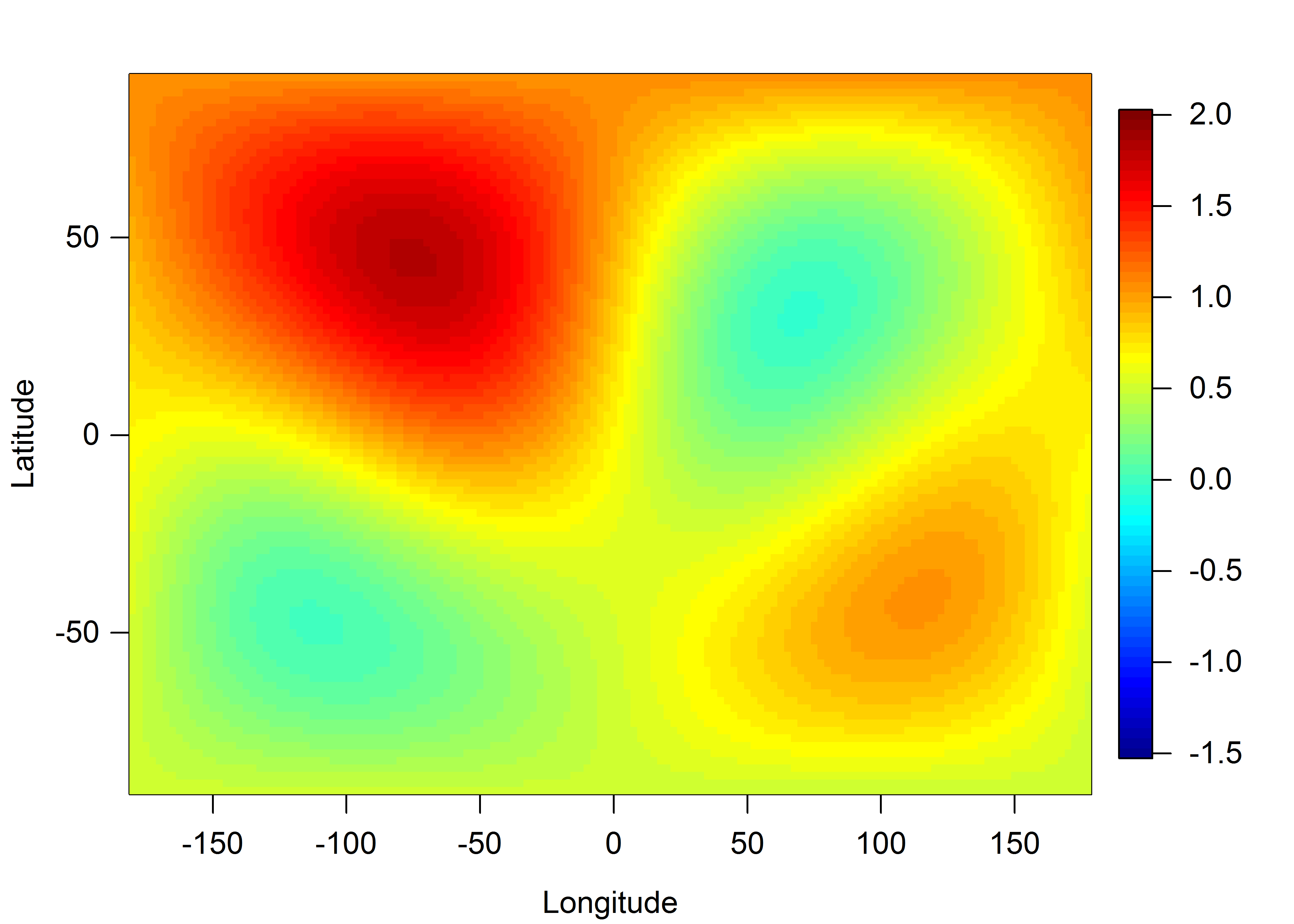}
        }\\          
  \end{center}
    \caption{The first 2 order expansion surface for $\kappa^F$ and $\tau^F$ for 
observations $y^F$.  }
    \label{fig:ex5-kappa}
\end{figure}

In order to understand the role of the SPDE parameters in the 
calibration, we then perform a calibration using only the  
coefficients $\{\kappa^M, \tau^M \}$.  Under the same priors and algorithm, the 
posterior mean and SD of the first three orders of the expansion for $\kappa^M$ 
and $\tau^M$ are shown in the second part of Table \ref{tab:ex5-post} (Strategies E - 
G).  Even though the calibration does not fully succeed (and should not without 
matching original outputs to observations but only SPDE information), the result 
in the 2nd order expansion for $\kappa^M$ and $\tau^M$ seems informative as the 
posterior modes are close to the true values. The first two orders of the 
expansion surface for $\kappa^F$ and $\tau^F$ for the observations are 
shown in Fig. \ref{fig:ex5-kappa}. It is difficult to directly interpret the 
features of $\kappa(s)$ and $\tau(s)$. However, from Fig. \ref{fig:ex5-kappa}(c)-(d) 
we can see that a strong northeast-southwest flow in $\kappa^F(s)$ matches 
the pattern in Fig. \ref{fig:ex5-surface} (b)-(c), and a highly 
anti-correlation between $\tau^F(s)$ (inverse precision) and $y^F$ surface. 

For the next step, we infer $\{c^M, \kappa^M, \tau^M \}$ jointly with the GP 
model.  We combined the coefficients in Strategies A-C (coefficients for the 
mean structure) and Strategies E-F (coefficients for the SPDE parameters).
The results are presented in the third part of Table \ref{tab:ex5-post}.  
We can see that with the SPDE information included, we achieve an improvement 
in the calibration by combining local structure with a global process. 
For example, Strategies C and K have a similar number of coefficients, but 
the combination with SPDE increases the estimated accuracy in $\theta_2$ and 
$\theta_3$. The similar case of Strategies D and L also supports the use of 
SPDE information. Strategy L uses a smaller number of coefficients, while 
achieving an improvement in terms of increased accuracy in $\theta_3$ and 
reduced RMSE. Nevertheless, only in the case of the 2nd order expansion do 
the SPDE parameters help; the 1st order expansion cannot achieve a good result 
in this example. This demonstrates that the nonstationarity is rather complex. 
From these findings we thus acknowledge that the SPDE technique enables us to 
identify the local feature from the global spatial process in the calibration. 
Therefore we highlight that when we cannot make an improvement in estimation 
accuracy by increasing the basis number into the mean structure, the SPDE 
technique can serve as a valuable alternative.

We provide more illustrations of the flexibility of our approach under various situations: (1) calibration with irregularly spaced outputs over the plane using B-spline basis ; (2) investigation of the connection between the calibration  accuracy and the number of computer runs $r$, and between the calibration accuracy and the orders and modes of SHs; (3) comparison of our approach and the empirical orthogonal functions (EOFs) approach and original \cite{KO01} framework, in the supplemental material for the interested reader.

\section{Application to the WACCM experiments}

A series of WACCM runs with the component set prescribed sea ice, data ocean and 
specified chemistry, with horizontal resolution $1.9\times 2.5^\circ$ and 66 
vertical levels were simulated from 1st January 2000. The GW parameterizations 
in WACCM depend on four inputs: (1) cbias ($\theta_1 \in [-5,5]$): anisotropy of 
the source spectrum, e.g. -5m/s: the spectrum has a stronger westward component, 
with center of the spectrum at 5m/s westward. Note that default simulation in 
WACCM is isotropic (i.e., cbias=0). An anisotropic GW source has been long reckoned 
to be a potential to improve the middle atmosphere circulation compared to an 
isotropic source \citep{medvedev1998role, hamilton2013gravity, chunchuzov2015study}; 
(2) effgw ($\theta_2 \in [0.05, 0.3]$): the efficiency factor, measures the gravity 
wave intermittency; (3) flatgw ($\theta_3 \in [1, 3]$): controls the momentum flux 
of the parameterized waves at the launch levels; (4) launlvl ($\theta_4 \in [50, 700]$): launch levels of the waves. The value of GW inputs $\boldsymbol\theta$ are generated 
by a maximin LHD (but scaled to be $[0,1]^4$). We simulated $r=100$ runs for 2 months. 
The first month was discarded as a spin--up period. Each output was computed over 96 latitudes and 144 longitudes, so the total output size is $n \times r=96 \times 144 
\times 100= 1382400$. We perform the calibration for the WACCM, either against 
synthetic (but with added nonstationary observation errors) or real observations, 
in order to fully validate our approach.

\subsection{Calibration against synthetic observations}
\subsubsection{Model set-up}
\begin{figure}[bht!]
    \begin{center}
        \subfigure[Standard output ($\eta^\ast$)]{%
        \includegraphics[height=4cm, width=0.5\textwidth]{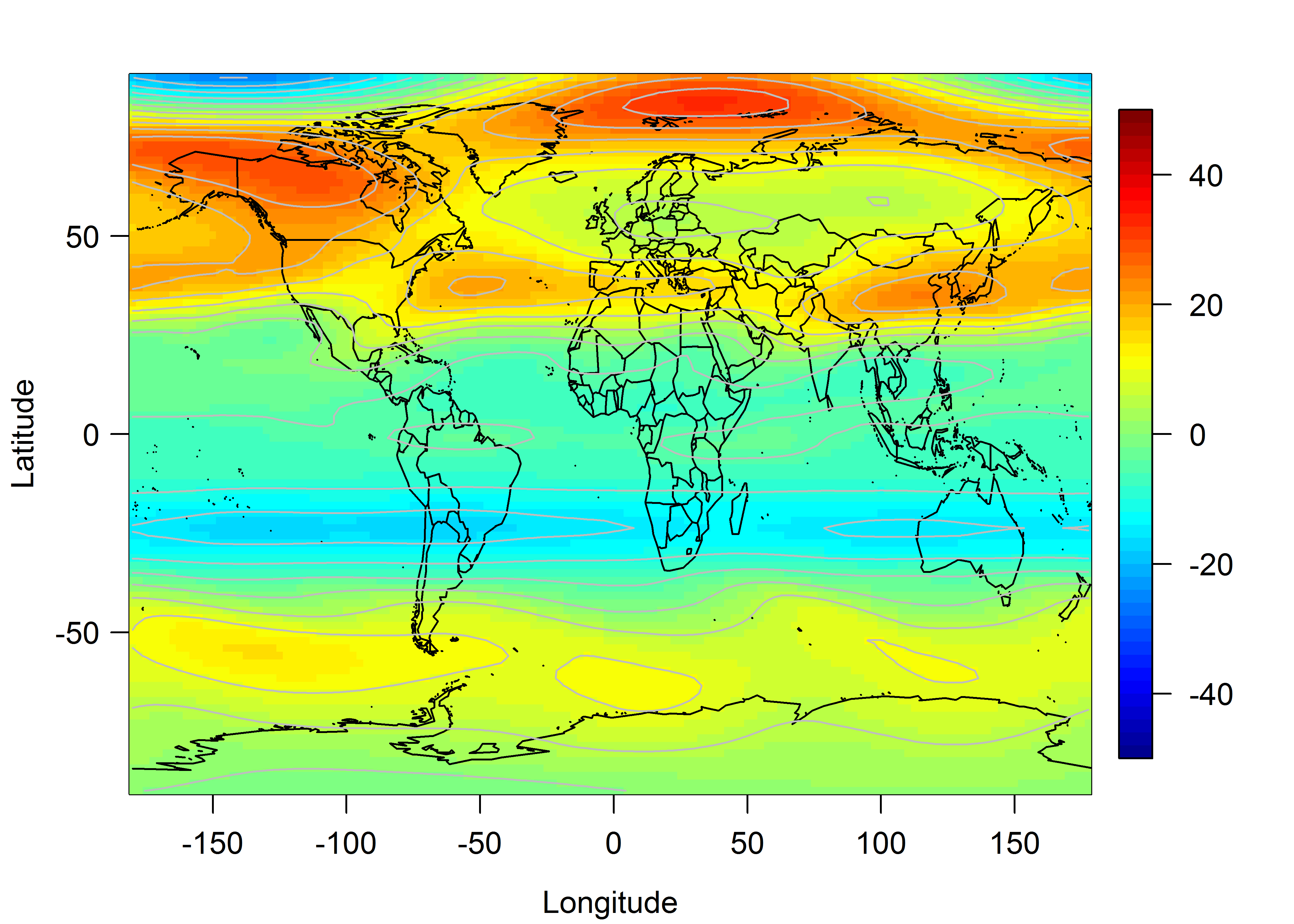}
        }%
        \subfigure[Standard output with noise ($y^F$)]{%
        \includegraphics[height=4cm, 
width=0.5\textwidth]{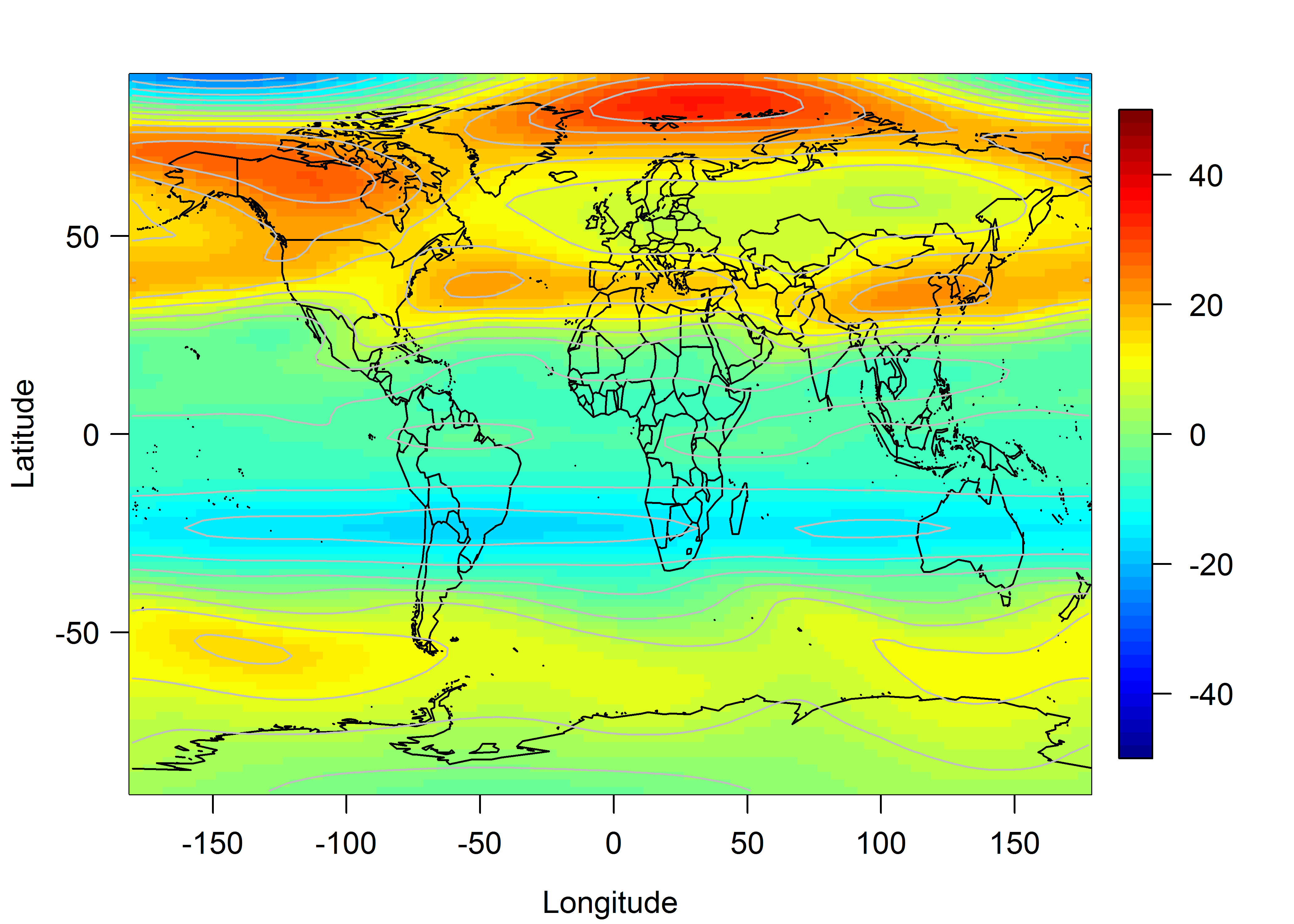}
        }\\          
        \end{center}
    \caption{(a) Zonal wind standard output; (b) Assumed observed surface: noise 
and discrepancy added to the zonal wind standard output (30mb, Feb. 2000).}
    \label{fig:30mb-uncertainties}
\end{figure}

To illustrate our methodology, we compare the zonal wind simulations 
$\eta(\mathbf{s}_i, \boldsymbol\theta_j)$, where $\mathbf{s}_i, i=1, \ldots, 
96\times 144$ are the latitude and longitude on the spherical domain, and 
$j=1,\ldots, 100$ is the index of the runs, with WACCM's standard outputs
(i.e. default simulation), instead of actual observations. Therefore we know the true GW parameters 
values and can validate our method. Let $\eta^\ast(\mathbf{s}_i)$ be the zonal 
wind surface from WACCM standard output. In order to account for possible 
observation error and lack of physics in the model (discrepancy), and thus evaluate the robustness of our method, we add a 
smooth noise to $\eta^\ast(\mathbf{s}_i)$ by assuming that the observations are given by:

\begin{align*}
y^F(\tilde{\mathbf{s}}_i) = \eta^\ast(\tilde{\mathbf{s}}_i) + 
\frac{\sigma_{\eta^\ast}}{5}s_1 + \frac{1}{2} s_2 s_3, 
\end{align*}
where $\tilde{\mathbf{s}}_i =(s_1, s_2, s_3)$ are the spherical coordinates, and 
$\sigma_{\eta^\ast} = 11.14$ is the SD of $\eta^\ast$.  Fig. 
\ref{fig:30mb-uncertainties}(a) and (b) shows the zonal wind surfaces from 
standard outputs and synthetic observations at 30mb, February 2000. 

As for the computational issue, in practice it is difficult to deal with a size of model 
output beyond moderately large (say $\simeq 2000$ responses). Here we have 
$r=100$ computer runs, therefore we seek to decompose each model output with 
about 20 coefficients.
We represent observations and model discrepancies using 3rd and 4th order SHT for model outputs and observations respectively. This allows enough flexibility.
We report two strategies (A) with or (B) without including 1st order SPDE nonstationary information. We also report two other strategies that use (C) 5 or (D) 10 principal components (with 95.8\% and 97.9\% respectively of the variation explained) to decompose the model outputs and observations (see algorithm in the supplemental material).

\begin{table}
\caption{\label{tab:post}Posterior mode of GW parameters on the rescaled $[0,1]$ range. \textit{Note that the MCMC did not converge in cases C and D, so these estimates are unreliable.}}
\centering
\begin{tabular}{l r r r r}
\hline
Strategy & cbias & effgw & flatgw & launlvl  \\
 & ($\theta^\ast_1 =0.5$) & ($\theta^\ast_2 =0.56$) & ($\theta^\ast_3 =0)$ & ($\theta^\ast_4 = 0.2308$) \\
\hline
A (SH-nonstationary SPDE) & \textbf{0.435} & \textbf{0.547} & \textbf{0.060} & \textbf{0.276}\\
B (SH-stationary SPDE)& 0.361 & \textbf{0.561} & 0.281 & \textbf{0.282}\\
C (5 PCs)& \textit{0.538} & \textit{0.396} &\textit{ 0.741} & \textit{0.523}\\
D (10 PCs) &\textit{ 0.639 }& \textit{0.082} & \textit{0.466} & \textit{0.908}\\
\hline
\end{tabular}
\end{table}

\subsubsection{Prediction accuracy}

The posterior modes of each strategy are shown in Table \ref{tab:post}. Both Strategies A and B calibrate $\theta_2$ well and slightly overestimate $\theta_4$. The inclusion of SPDE parameters in A v. B not only increases the accuracy of the posterior mode for $\theta_1$, but estimate very closely $\theta_3$, a difficult task as the true value lies on the lower bound. The quantification of the anisotropic velocity in a large spatial process is a difficult problem \citep{large2001equatorial, lauritzen2015ncar}. The improvement of accuracy in the estimation of $\theta_1$ confirms the value of using the SPDE technique in the calibration since the nonstationarity allows the amount of flexibility required to identify much clearly the value of $\theta_1$. 

Unfortunately, the MCMCs do not converge for strategies C and D, hence their posterior modes are uninterpretable (but we report them nevertheless). Note that this result can be expected. Our GW parameterization aims to reduce zonal wind bias at the Tropics associated with the QBO; however, the principal mode of variability in our model outputs occurs across the Northern Hemisphere (in East Asia to be precisely), where the influence of the GW is indirect (see next section for further discussion in comparison with real observations). The PC decomposition will focus on the variability in Northern Hemisphere compared to the Tropics. Recent studies suggest that PC-based approach tends to cause a `terminal case analysis' in the climate modeling \citep{salter2018uncertainty}, which means that there is no set of parameters that can allow the model to mimic reality. For this reason the PC-based approach is not appropriate for our calibration setting.

\begin{figure}[bht]
    \begin{center}
        \subfigure[2nd SHT (9 coefficients)]{%
       \includegraphics[height=4cm, width=0.5\textwidth]{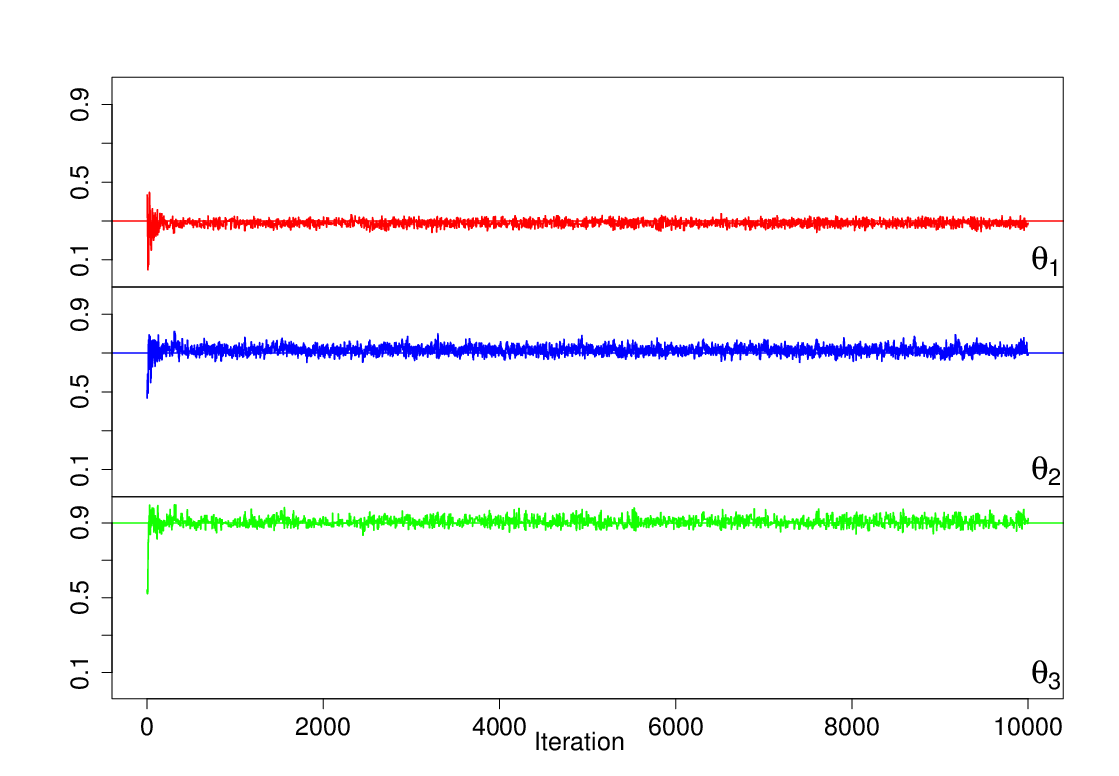}
        }%
        \subfigure[9 EOFs representation]{
        \includegraphics[height=4cm, width=0.5\textwidth]{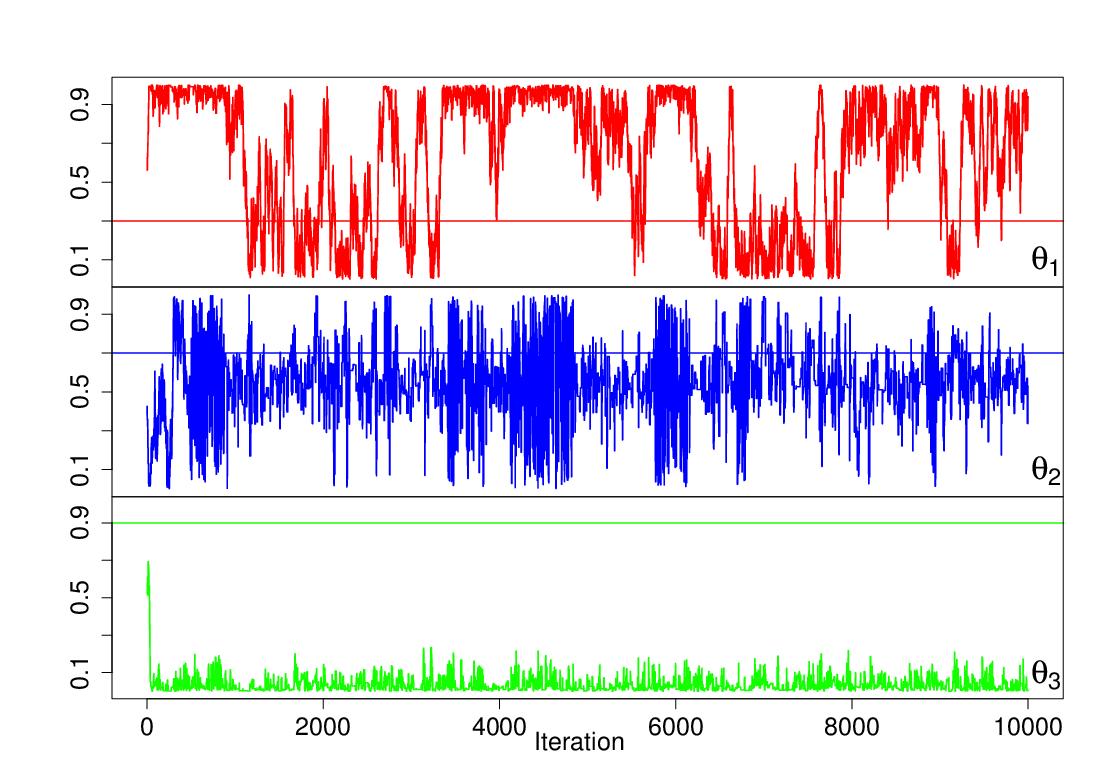}
        }\\
    \end{center}
    \caption{MCMC paths of $\boldsymbol\theta$ for our approach and PC-based algorithms (synthetic case in supplementary material). Solid lines indicate the true values. }
    \label{fig:ex3-mcmc}
\end{figure}

Fig. \ref{fig:ex3-mcmc} shows a concrete example of such lack of convergence in a synthetic example (see supplementary material). This is a comparison of the MCMC sample paths of the calibration parameters by 2nd order SHs (9 coefficients) and 9 PCs representation.  We can see that for all calibration parameters in the SHs approach, convergence occurred after roughly 500 iterations, whereas chains do not converge in the PC approach.

\begin{figure}[h!]
    \begin{center}
        \subfigure[Model A ; $\boldsymbol\rho_\eta$]{%
        \includegraphics[width=0.65\textwidth]{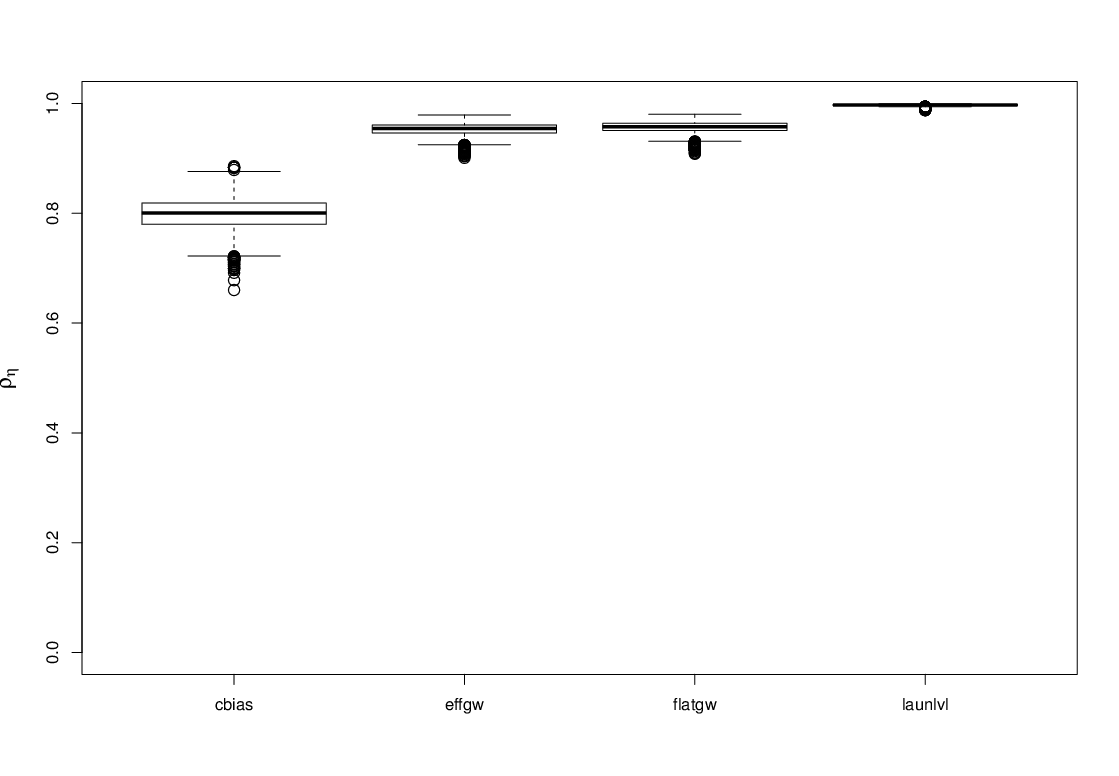}
        }%
        \subfigure[Model A: $\boldsymbol\theta$]{%
        \includegraphics[width=0.35\textwidth]{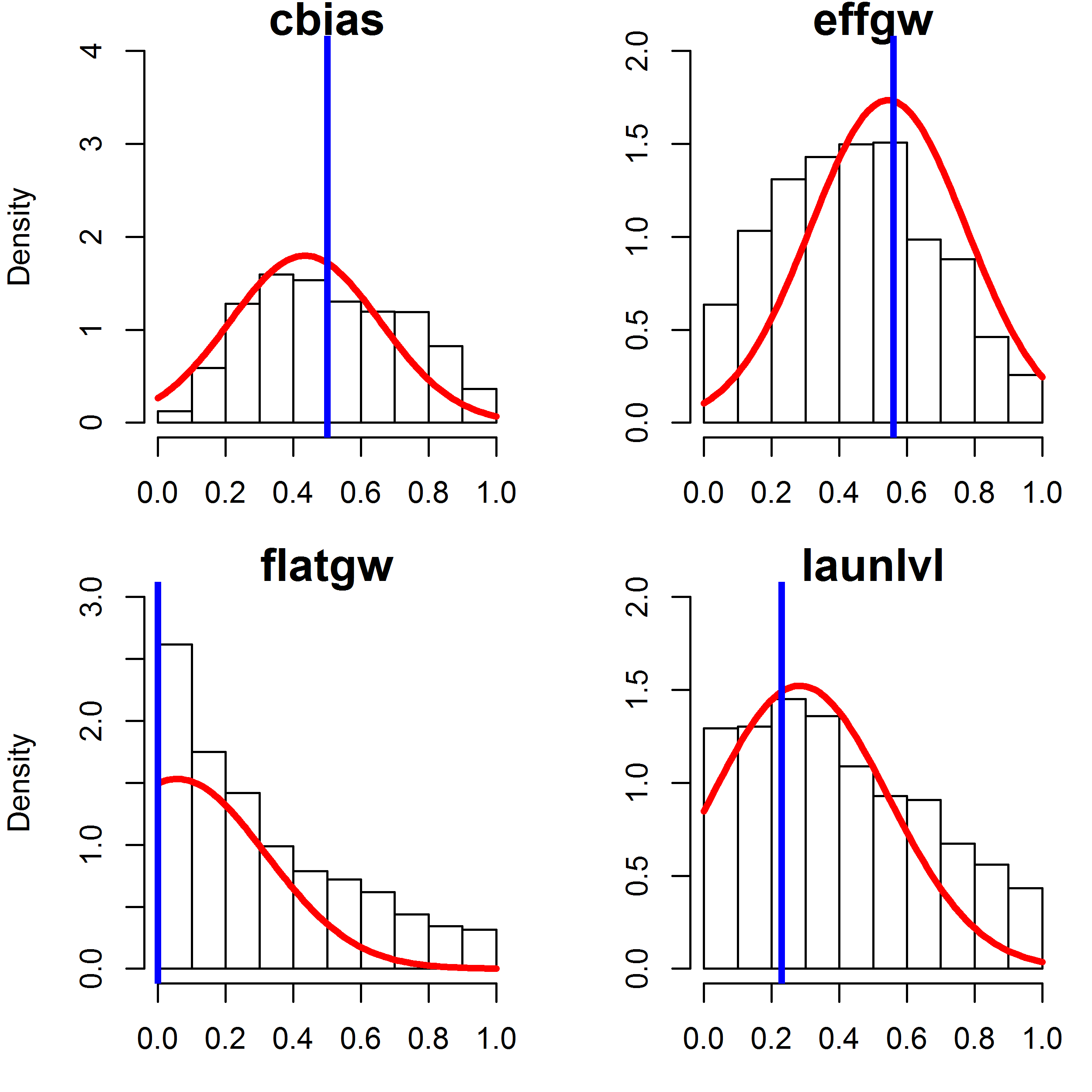}
        }\\           
        \subfigure[Model B: $\boldsymbol\rho_\eta$]{%
        \includegraphics[width=0.65\textwidth]{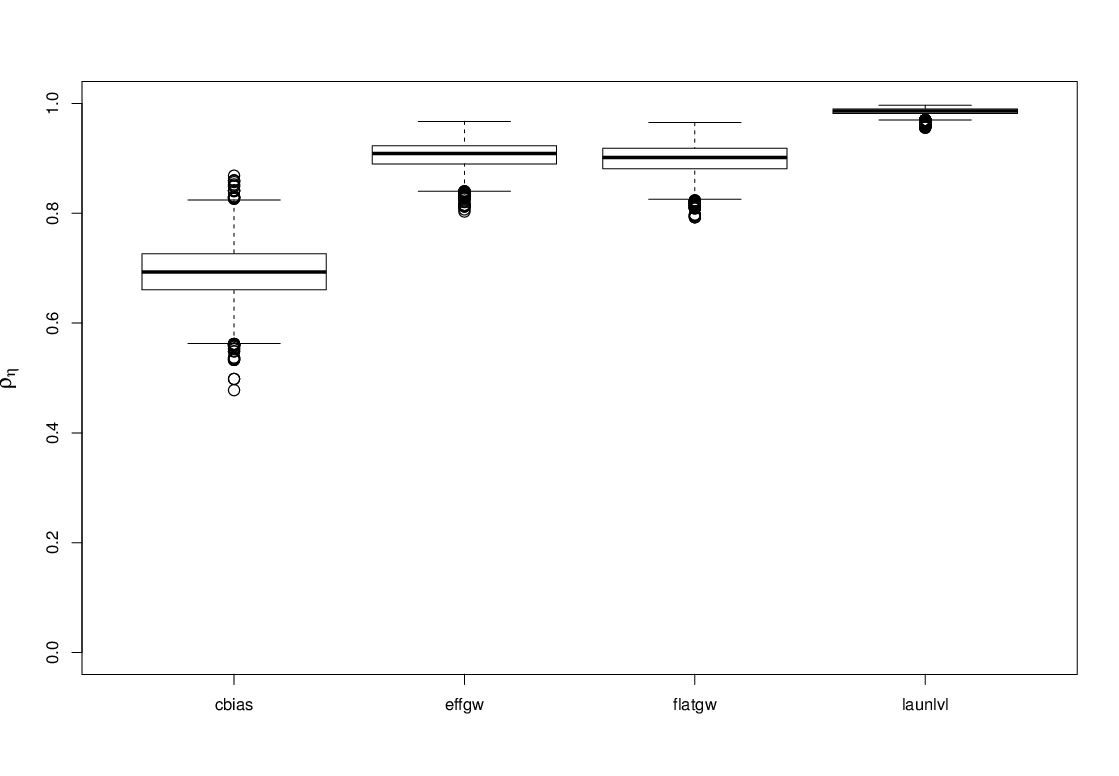}
        }%
        \subfigure[Model B: $\boldsymbol\theta$]{%
        \includegraphics[width=0.35\textwidth]{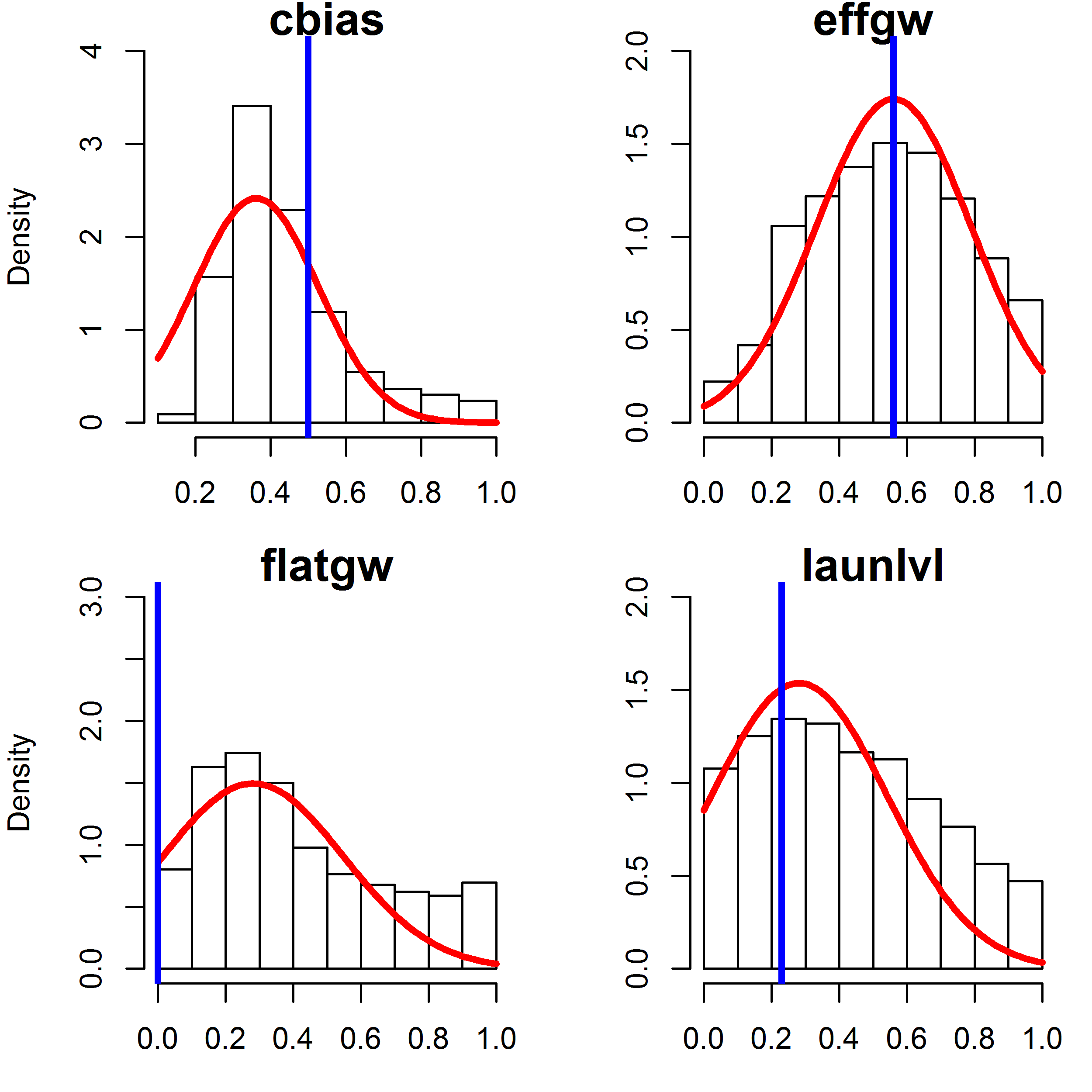}
        }\\           
     \end{center}
    \caption{Boxplots of the marginal posterior distribution for correlation 
parameters $\boldsymbol\rho_\eta$ (left panels); and marginals for the posterior distribution 
of the GW parameters $\boldsymbol\theta$ (right panels). Vertical lines indicate the true 
values (30mb, Feb. 2000).
    }
    \label{fig:30mb-post}
\end{figure}

The left panels of Fig. \ref{fig:30mb-post} show the boxplots of the 
marginal posterior distributions for the $\rho_\eta$'s for Strategies 
A and B, which control the dependence strength in each pair of $\boldsymbol\theta$ 
in the GP model. The posterior density of $\rho_4$ closes to 1 that indicate a 
very weakly significant effect for $\theta_4$. The marginal posterior 
densities for each $\boldsymbol\theta$ are displayed in the right panels of 
Fig. \ref{fig:30mb-post}. Our approach provides a good compromise 
between computational feasibility and fidelity to the data by only using 
parsimonious representations. The results suggest that our technique 
on calibration of global-scale outputs is effective.

\subsection{Calibration against real observations}

\subsubsection{Posterior sampling}
\begin{figure}[ht!]
    \begin{center}
        \subfigure[ERA obs.]{%
        \includegraphics[height=4cm, width=0.5\textwidth]{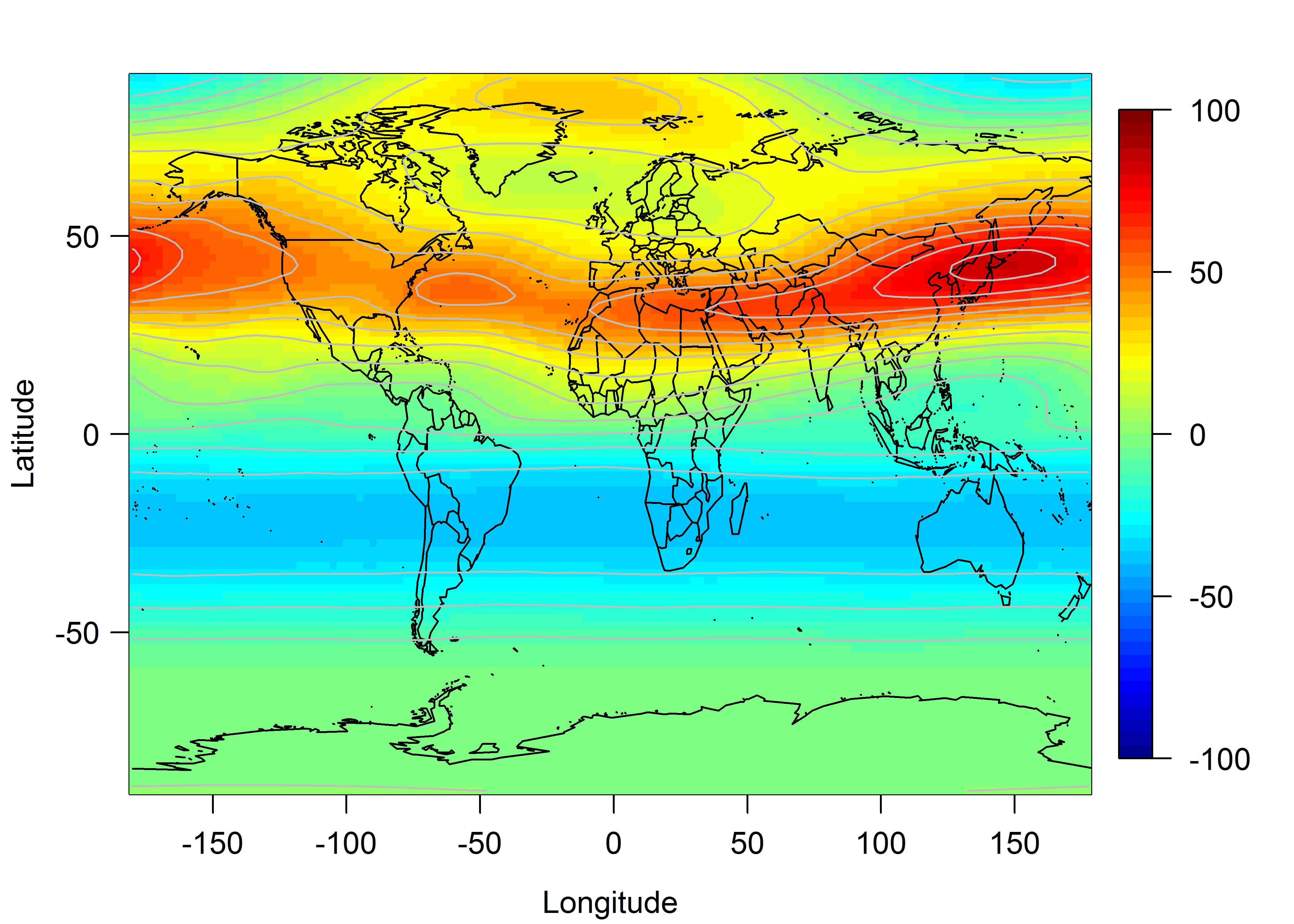}
        }%
        \subfigure[Standard output]{%
            \includegraphics[height=4cm, width=0.5\textwidth]{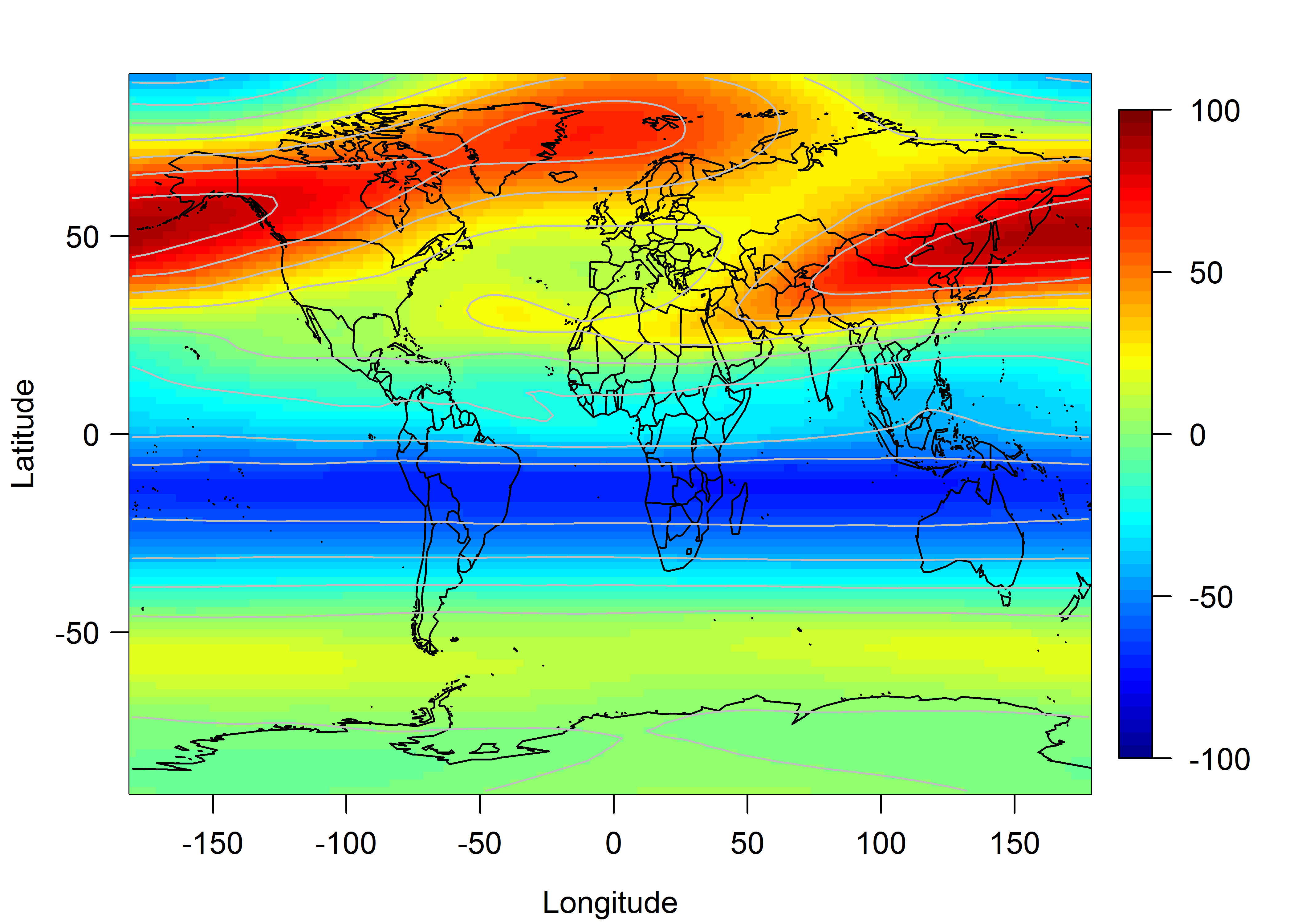}
        }\\       
        \subfigure[MCMC paths]{%
            \includegraphics[height=4cm, width=0.5\textwidth]{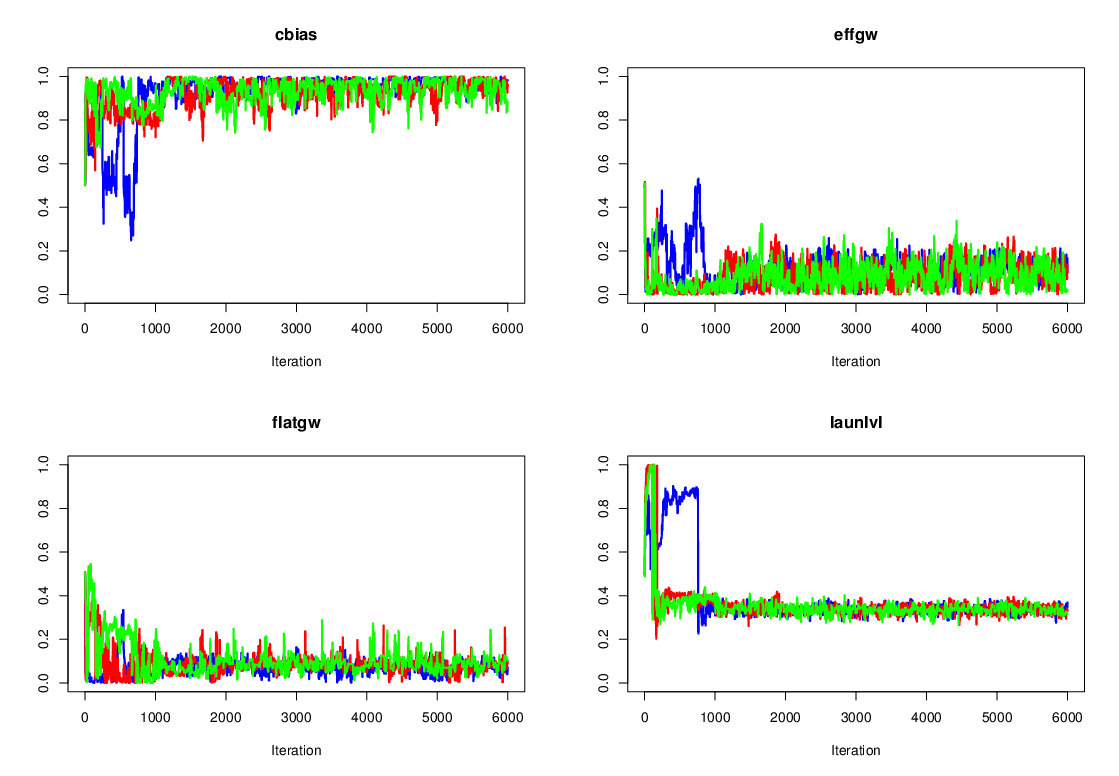}
        }%
        \subfigure[Calibrated output]{%
            \includegraphics[height=4cm, width=0.5\textwidth]{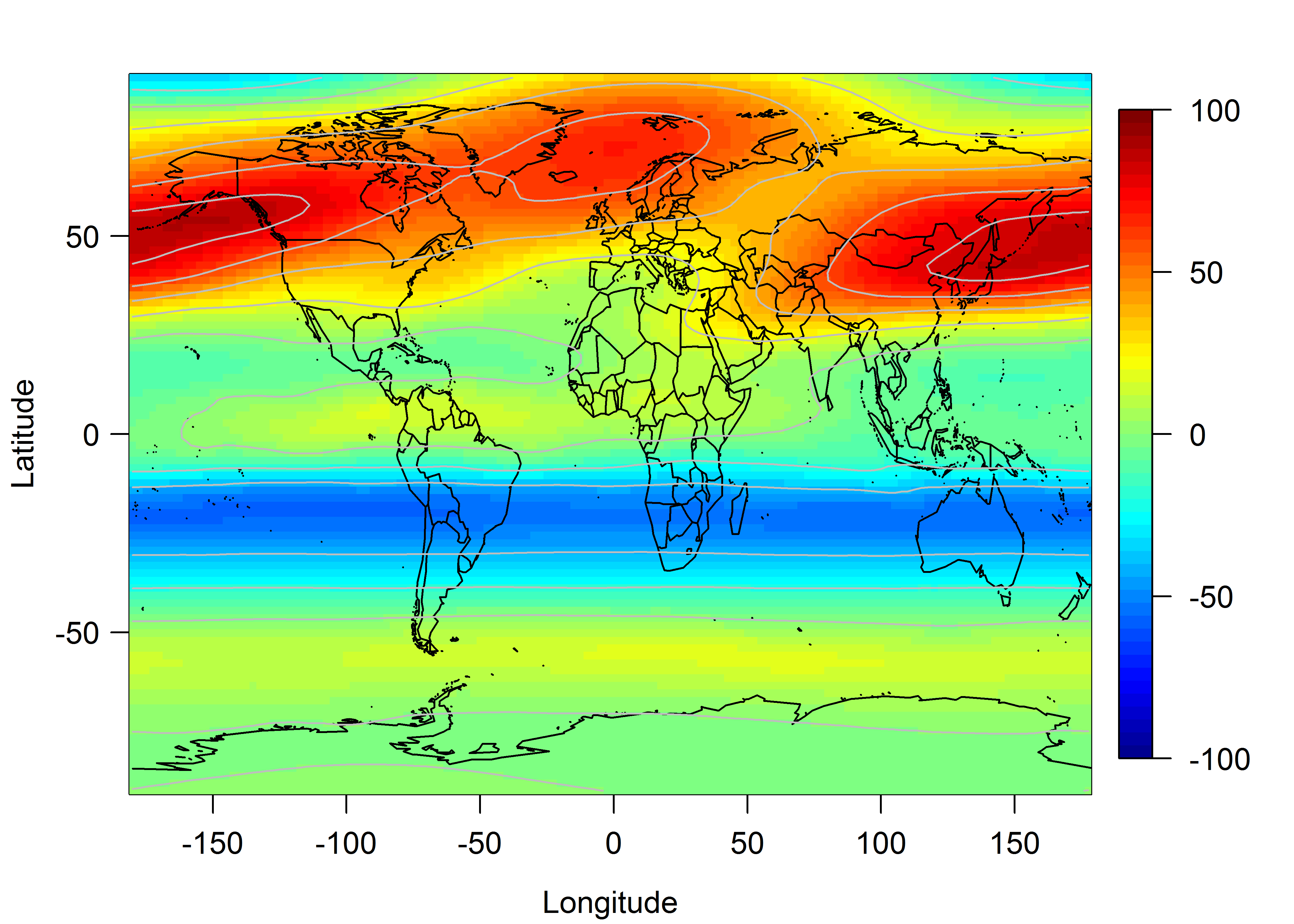}
        }\\ 
\end{center}
    \caption{Zonal wind from the (a) ERA data and (b) WACCM standard output;  (c) MCMC paths for 3 chains and (d) Zonal wind generated by posterior from the calibration (1mb, Feb. 2000).}
    \label{fig:1mb-output}
\end{figure}

The final step is to carry out the calibration against real observations. We use 
zonal wind data obtained from the \textit{European Centre for Medium-Range 
Weather Forecasts} (ECMWF) 40 Year Re-analysis (ERA) Data Archive.  We focus on 
the altitude of 1mb, as the outputs in low altitudes are less sensitive to GW 
parametrizations and match the observations already well. Fig. 
\ref{fig:1mb-output}(a) and (b) shows the ERA observations and zonal wind 
surfaces from standard outputs at 1mb, February 2000. 
Under the same settings described in the previous section, Fig. 
\ref{fig:1mb-output}(c) shows the MCMC paths for 3 chains, with 6000 
iterations, corresponding respectively to the calibration parameters. The 
convergence of the MCMC chain can be established for the parameters $\theta_2$, 
$\theta_3$ and $\theta_4$, with posterior modes 0.107 (SD=0.051), 0.081 
(SD=0.029) and 0.339 (SD=0.018) in the $[0,1]$ scale, respectively. 
The posterior mode of $\theta_1$ lies in upper bound. 
 We then use posterior modes for these paths, collected as input values for the 
validation of WACCM. The calibrated output displayed in Fig. 
\ref{fig:1mb-output}(d), shown a root-mean-square error (RMSE) of 18.15, 
which is a percentage of improvement of 14.99\% over the standard output (the 
RMSE between ERA observations and standard output is 21.35).

The resulting histograms for the calibration parameters, with first 1000 iterations dropped as they are reckoned to be in burn-in, are shown in Fig. 7. As expected from MCMC plot, a normal distribution can be established for $\theta_2$, $\theta_3$ and $\theta_4$. The distribution of $\theta_1$ shows is skewed against the upper bound. It means that the possible calibrated value may lie outside the boundary. Since $\theta_1$ represents the anisotropic velocity of zonal wind (the model default is assumed to be isotropic), the results suggest that we would need a more eastward component. It seems that this is a spurious effect of the simplicity in the parameterization. Indeed, in order to avoid losing the westward components and to acknowledge the physical reality, it may be helpful to have a ``bi-modal" spectrum \citep{arfeuille2013modeling, zhu2017development}, with one peak in the eastward direction and another in the westward, and these two components do not have to be the same. Indeed, uneven amplitudes of the QBO easterly and westerly phases are often observed in the previous studies \citep{naujokat1986update, garcia1997climatology, ern2008equatorial}.  GWs schemes are currently under development within the NCAR WACCM working group to improve the representation of the QBO and help fix the cold pole problem \citep{garcia2017modification}. Further development will allow us to have more flexible GWs schemes, but it is beyond the scope of the present setting for the climate simulation.

\begin{figure}[ht!]
    \begin{center}
        \includegraphics[width=0.7\textwidth]{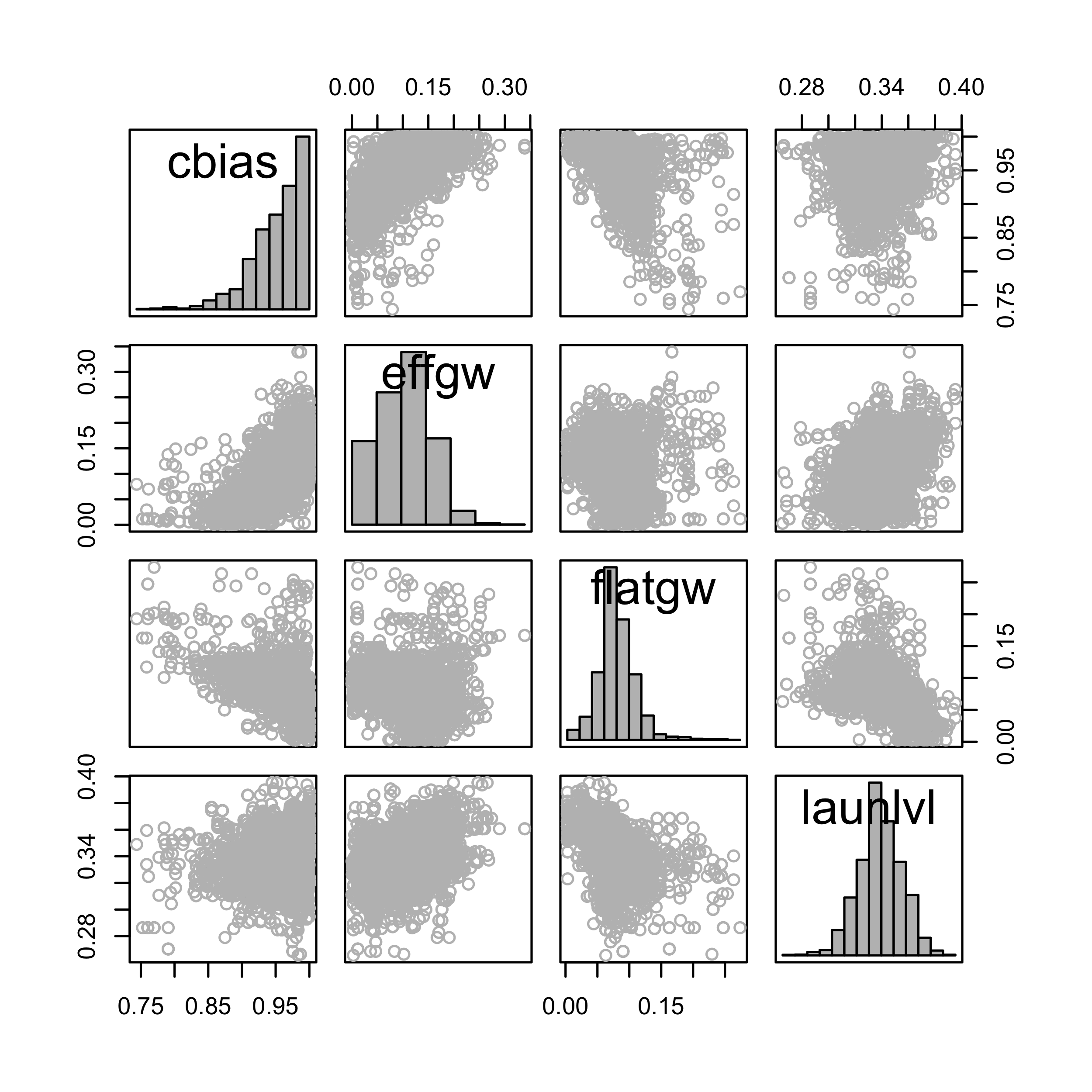}
    \end{center}
    \caption{Density of posterior calibration parameter for zonal wind simulation (1mb, Feb. 2000).}
    \label{fig:1mb-post} 
\end{figure}

\begin{figure}[ht!]
    \begin{center}
        \subfigure[Zonal means]{%
        \includegraphics[height=4cm, width=0.5\textwidth]{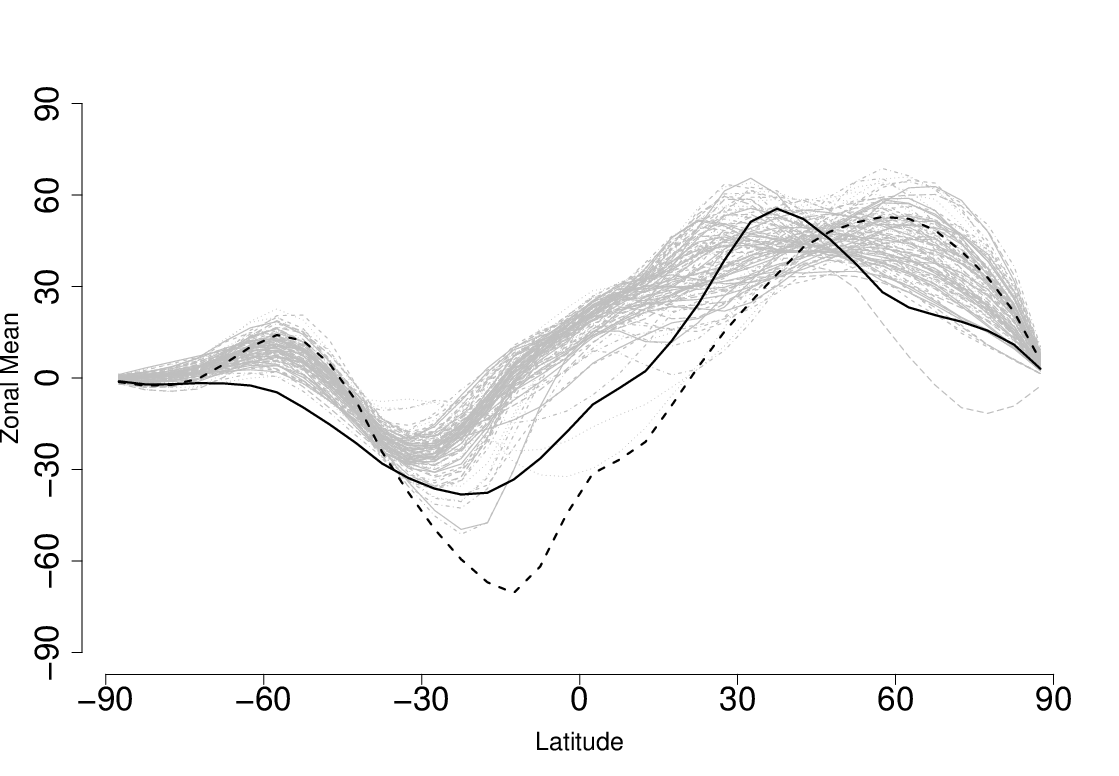}
        }%
        \subfigure[Grid-by-grid SDs]{%
        \includegraphics[height=4cm, width=0.5\textwidth]{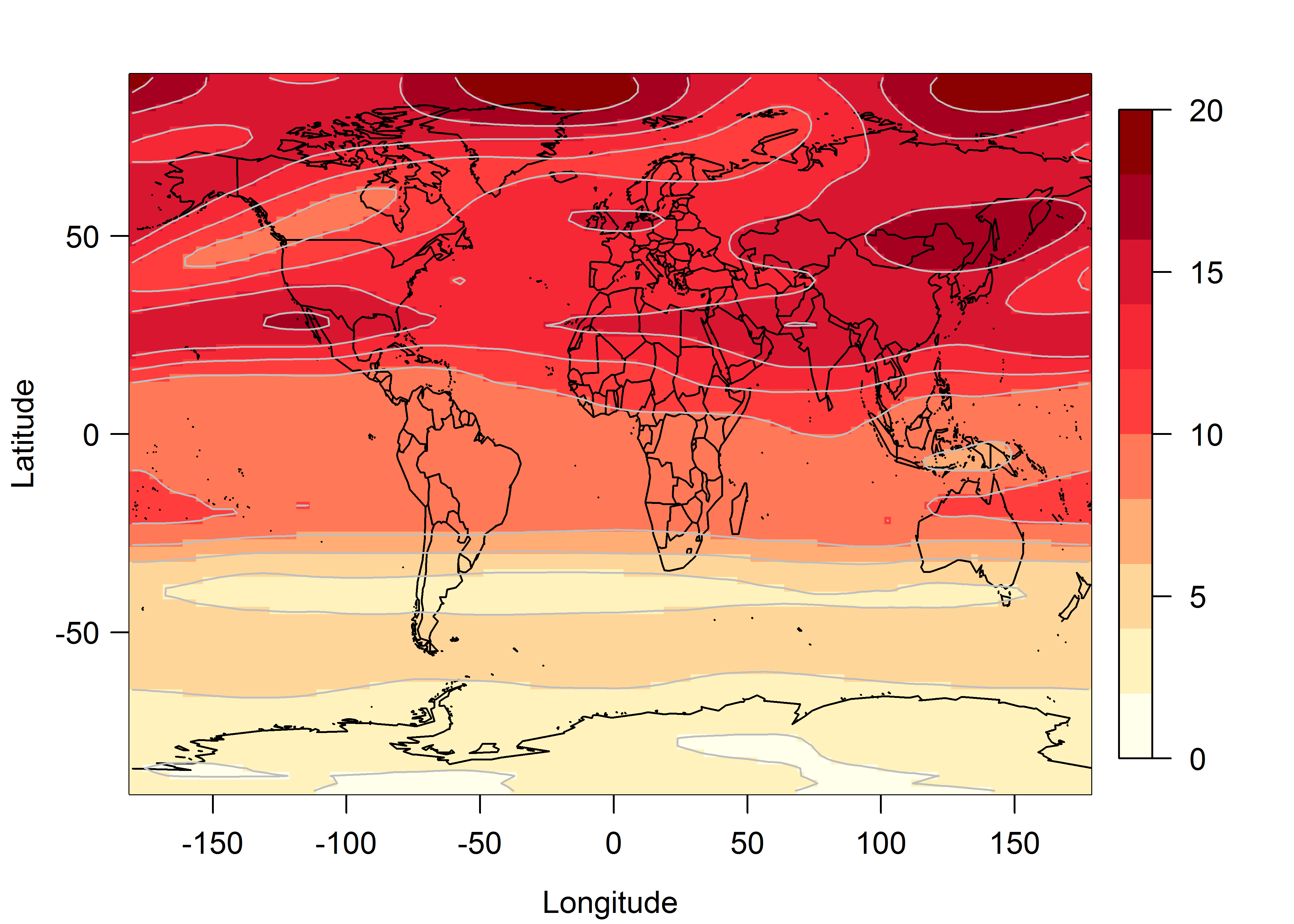}
        }\\ 
        \subfigure[Grid-by-grid differences]{%
            \includegraphics[height=4cm, width=0.5\textwidth]{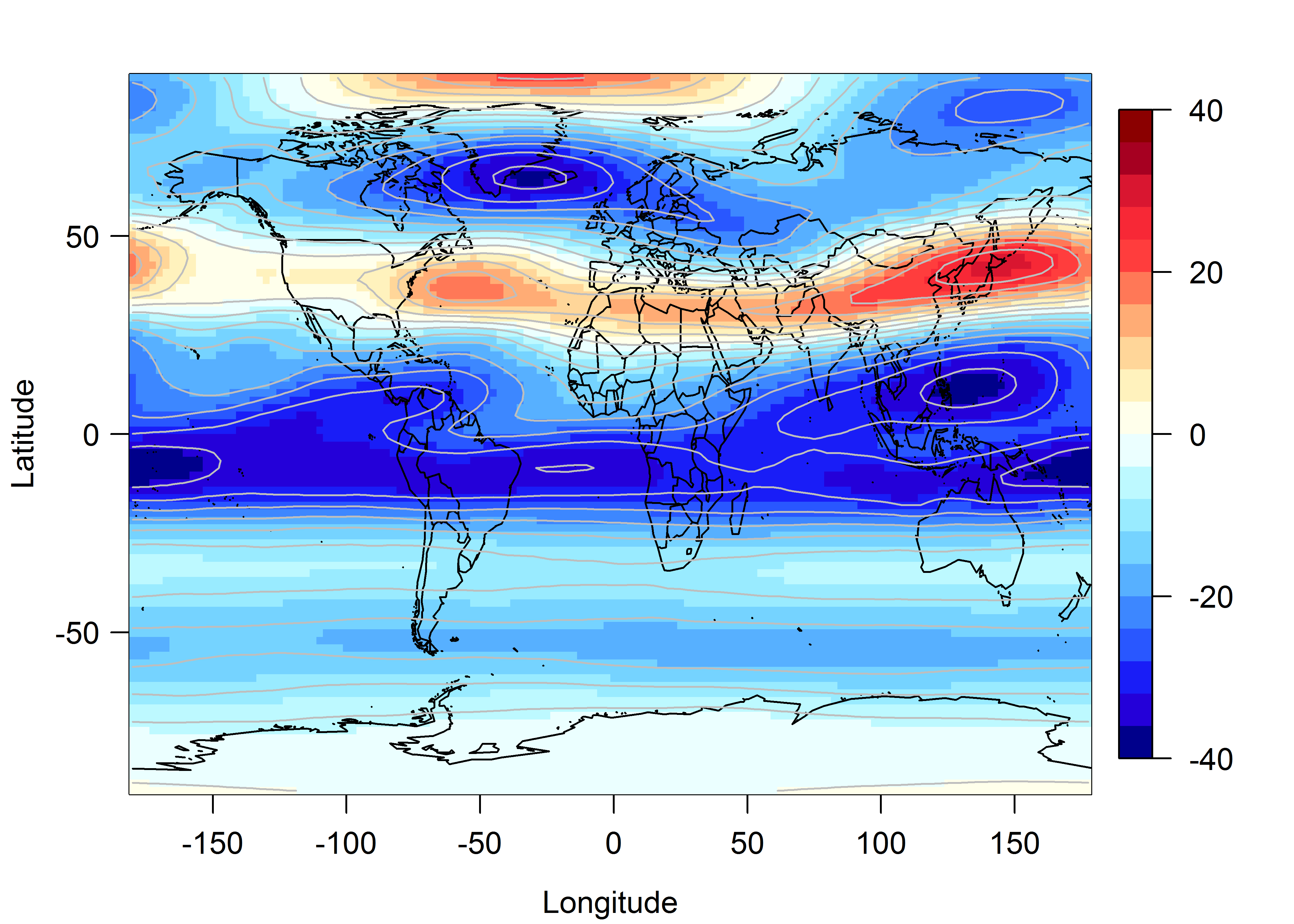}
        }%
        \subfigure[Model mean discrepancies]{%
            \includegraphics[height=4cm, width=0.5\textwidth]{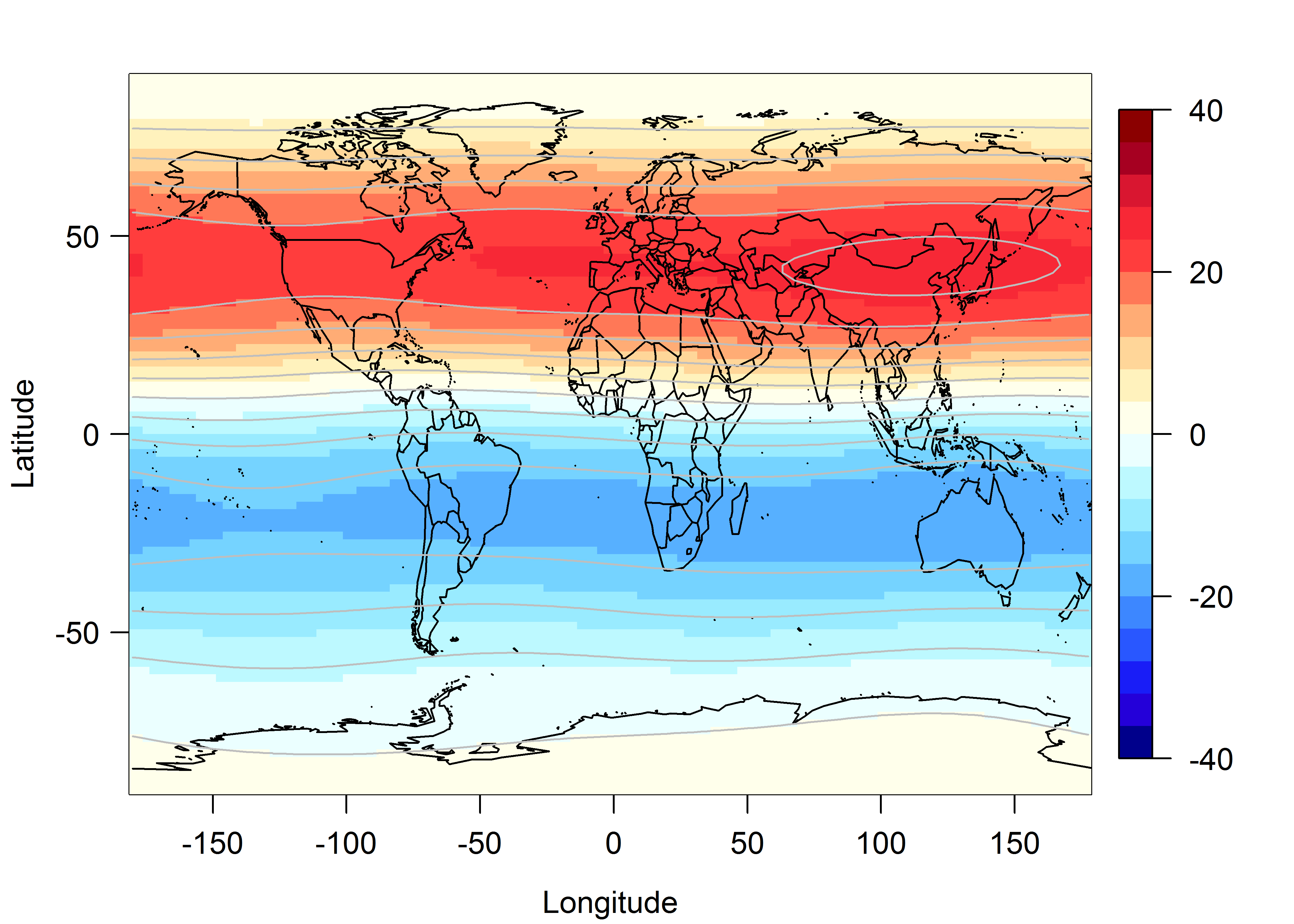}
        }\\
 \end{center}
    \caption{(a) Zonal means of observations (black solid line), standard (black dashed line) and model outputs (grey dotted lines); (b) Grid-by-Grid SDs map across model runs; (c) Differences between observations and mean structure of model outputs and (d) Model mean discrepancies map (1mb, Feb. 2000).}
    \label{fig:1mb-validation} 
\end{figure}

\subsubsection{Model discrepancy and uncertainty}
In order to assess the model uncertainty, Fig. \ref{fig:1mb-validation}(a) shows the zonal means calculated over every $5^\circ$ 
belt of observations (black solid line), standard outputs (black dashed line) 
and each run of model output (grey dotted lines): note that the zonal means over the Tropics are high compared to the 
observations and standard outputs. The input value of $\theta_3$ in the standard 
output is being at the lower border of parameter range, this may produce 
relatively extreme behavior over the Tropics in our model runs. Fig. 
\ref{fig:1mb-validation}(b) represents the grid-by-grid SDs map across model 
outputs. We can see that the spatial process is clearly anisotropic and highly 
latitude dependent; the uncertainties are concentrated over the Northern 
Hemisphere, and little significant variabilities can be found over the Southern 
Hemisphere. 
Fig. \ref{fig:1mb-validation}(c) compares the differences between observations 
and mean structure of model outputs in each cell (with respect to 100 LHD), 
i.e. $\delta_{\text{initial}}(\mathbf{s})=y^F(\mathbf{s}) - 
\bar{\eta}(\mathbf{s}, \boldsymbol\theta)$, where $\bar{\eta}(\mathbf{s})$ is 
the output means over space. This figure provides potential features of model 
discrepancy over space (albeit not the true discrepancy). As expected, the model 
tends to overestimate the values over the Tropics, which matches the pattern in 
Fig. \ref{fig:1mb-validation}(a).
Besides, this surface seems to match the pattern in Fig. 
\ref{fig:1mb-validation}(b). The largest model bias (apart from the Tropics) and 
variability both occur over Northeast Asia and the North Pole. Fig. 
\ref{fig:1mb-validation}(d) shows the posterior mean discrepancies surface in 
the sense of $\delta^\ast (\mathbf{s})= y^R(\mathbf{s}) - \eta(\mathbf{s}, 
\boldsymbol\theta^\ast)$. Our calibration reduces the bias (i.e. overestimation) 
over the Tropics, as well as the bias (i.e. underestimation) over the North 
Pole, whereas the bias over the Northeast Asia remains.

\subsubsection{Validation}
\begin{figure}[!]
    \begin{center}
        \subfigure[Global RMSEs]{%
            \includegraphics[height=4cm, 
width=0.5\textwidth]{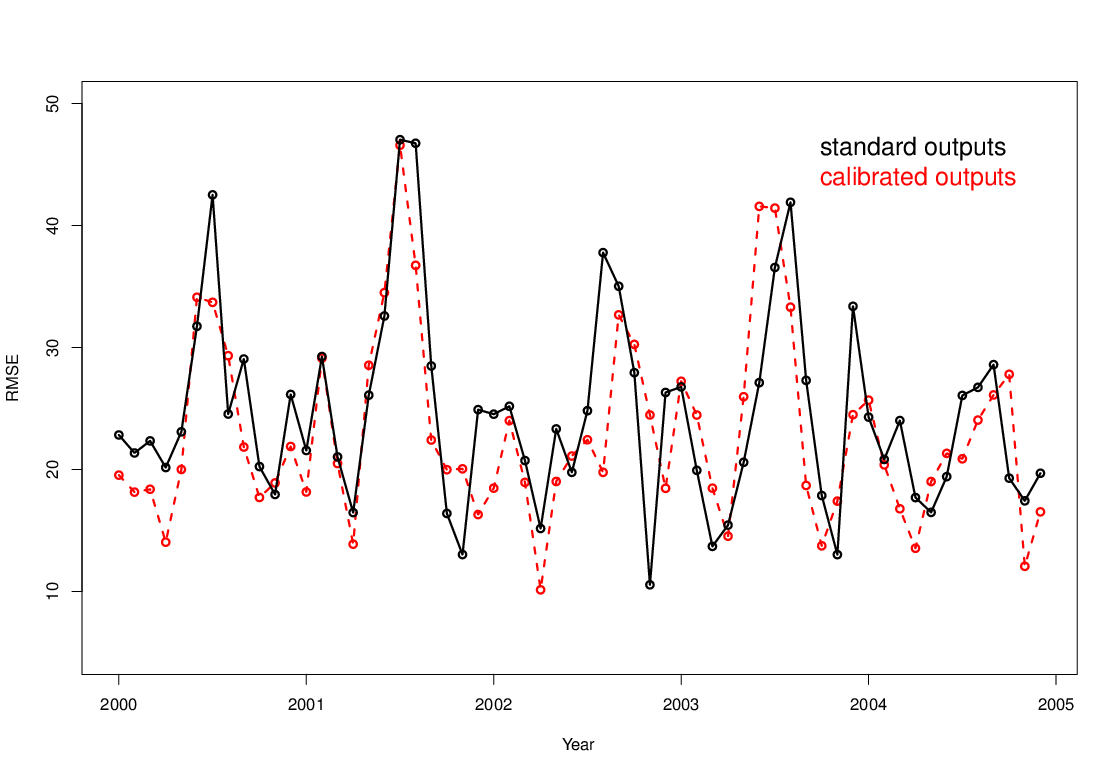}
        }%
        \subfigure[Tropical RMSEs]{%
            \includegraphics[height=4cm, 
width=0.5\textwidth]{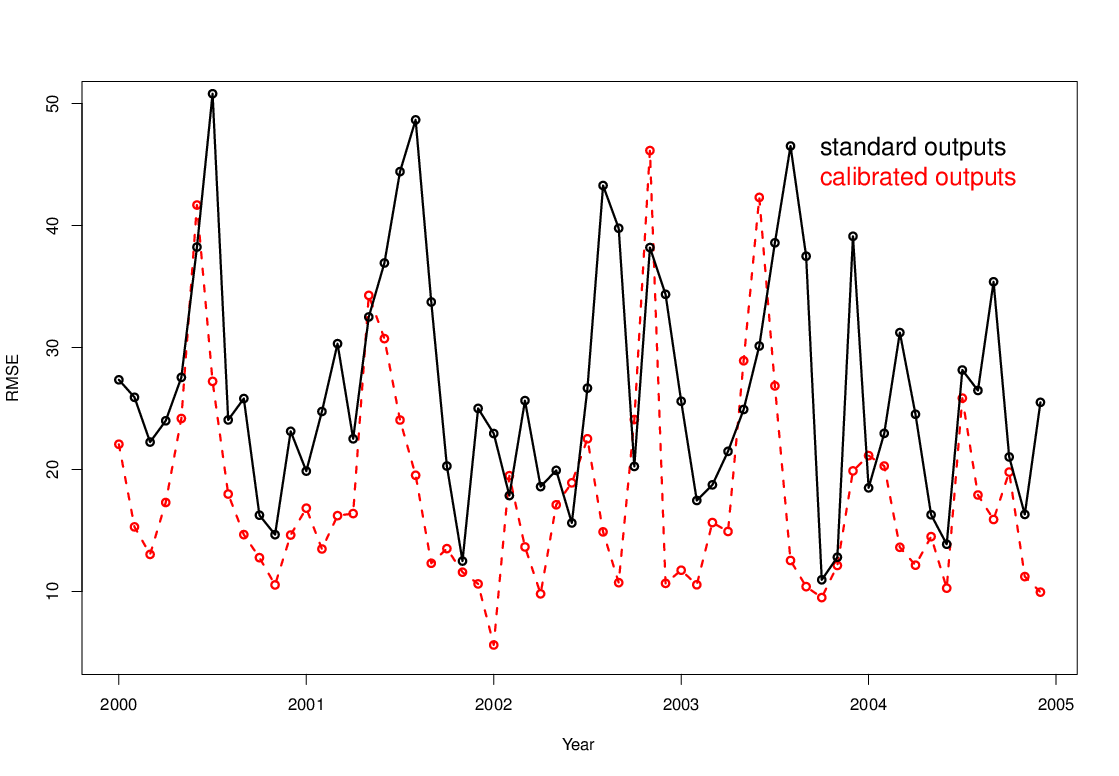}
        }\\        
  \end{center} 
    \caption{Monthly RMSE trends between the ERA observations and standard 
outputs (solid line) or calibrated outputs (dashed line), from 2000 to 2004.}
    \label{fig:rmse-val-1mb}
\end{figure}

We use the mode from each posterior distribution to simulate 5 years (2 QBO 
cycles) of zonal wind output. Fig. \ref{fig:rmse-val-1mb}(a) shows monthly RMSEs 
at 1mb globally, from 2000 to 2004. The overall averaged RMSE for the standard 
and calibrated outputs are 24.51 and 22.99, respectively, a small improvement. 
Indeed, our inertial GW scheme is designed to reduce the zonal wind bias over 
Tropics, we should not expect that our calibration will improve model simulations globally.
We thus investigate RMSEs over the Tropics over the same period. The 
RMSE trends are shown in Fig. \ref{fig:rmse-val-1mb}(b). The overall averaged 
RMSE over the Tropics for the standard and calibrated outputs are 26.64 and 
17.87, respectively. Therefore the improvement is more significant over the 
Tropics, with percentage of improvement 32.9\%. Simulations by our calibrated 
outputs outperform the standard code in 51 months out of 60 months. The 
calibration of WACCM with real observations over the whole output domain 
(i.e. including across altitudes) constitutes another level of complexity that 
needs joint scientific and statistical expertise. It is currently under investigation, 
but is beyond the scope of this paper. Indeed, observations are scarce at these 
altitudes and show features that require specific understanding of the upper 
atmosphere dynamics before being used for calibration, and over many years of 
simulation for an adequate comparison.

\section{Conclusion and discussion}
\label{sec:conc}
Our approach improved the calibration of large--scale computer model outputs 
distributed over spaces, parsimoniously, by using bases representations for 
the mean structures of the spatial surfaces. In addition, the INLA--SPDE 
approach was used to decompose its parameters characterizing nonstationarity 
over the the same bases in order to improve calibration. The synthetic and 
real examples confirm the ability of our approach to efficiently and accurately 
perform calibration. Our method was inspired by the wavelets method of 
\cite{bayarri2007computer}, but with a different type of outputs: spatial v. 
time series.  We can expect that the spherical wavelet decomposition may also 
be a possible alternative basis representation on the spatial domain, whenever 
appropriate (e.g. for sharp variations).

Another advantage of using the SH basis, compared to data-driven ones such as 
PCs, is that sequential design is allowed \citep{beck2016sequential}, because 
the basis elements will not change, and model runs are obtained at the same grids 
or scattered locations. In this study we illustrate our technique to a specific 
horizontal output from the WACCM simulator. The SHT of model outputs can also be 
extended to time varying processes. As noted by \cite{Jones63}, if a random 
field on a sphere varies with time, the representation becomes
$\eta(\mathbf{s},t)=\sum_{k=0}^\infty \sum_{h=-k}^k 
c_{k,h}(t)\psi_{k,h}(\mathbf{s})$,
where $ c_{k,h}(t)$ being an ordinary one-dimensional stochastic process. The 
set of all $ c_{k,h}(t)$ form an infinite dimensional stochastic process. 
Theoretically we can represent model outputs in space--time settings with such 
representations. Nevertheless, in climate or chemistry--transport simulations, 
we often encounter not only outputs in time and horizontal resolution, but also 
in vertical resolution. Therefore extensions to 4 dimensional correlations are 
needed, but they must maintain the computational tractability.

In our approach the covariance matrix is formulated as a block diagonal 
structure. We could relax this assumption and then adopt the block composite 
likelihood approach to accelerate the algorithm \citep{chang2015composite}. 
Unfortunately, this approach only cover the stationary case (though could be 
extended). Our approach naturally and efficiently models nonstationarity in 
space. Furthermore, there are cases where our approach is computationally more 
efficient than \cite{chang2015composite}. Indeed, if $m$ is large, their 
computational cost is about $O(\sum_{i=1}^B m_i^3)$, where $\sum_{i=1}^B m_i =m$ 
(depends on number and size of blocks $m_i$), whereas our cost is $O(N^3_y r^3)$, 
which is lower in many, but not all, applications. Since our climate experiment 
involves direct input-output projection, another potential extension of our 
approach is to combine recent development on Bayesian treed calibration technique, 
which partitions input space into subregions where our reduced rank approach can be 
applied, to further accelerate the calibration \citep{karagiannis2017bayesian, konomi2017bayesian}.

\bibliographystyle{apa}
{\footnotesize \bibliography{ref2}}
\end{document}